\documentclass[conference]{IEEEtran}
\IEEEoverridecommandlockouts
\usepackage{cite}
\usepackage{amsmath,amssymb,amsfonts}
\usepackage{graphicx}
\usepackage{textcomp}
\usepackage{xcolor}
\def\BibTeX{{\rm B\kern-.05em{\sc i\kern-.025em b}\kern-.08em
    T\kern-.1667em\lower.7ex\hbox{E}\kern-.125emX}}

\usepackage{subfigure}
\usepackage{enumerate}
\usepackage[ruled]{algorithm2e}
\usepackage{amsthm}
\usepackage{booktabs}
\usepackage{color}
\usepackage{multirow}

\newcommand{\change}[1]{#1}
\newcommand{\todo}[1]{#1}
\newcommand{\review}[1]{#1}
\newtheorem{definition}{Definition}
\usepackage{etoolbox}
\makeatletter
\patchcmd{\@makecaption}
  {\scshape}
  {}
  {}
  {}
\makeatletter
\patchcmd{\@makecaption}
  {\\}
  {.\ }
  {}
  {}
\makeatother

\begin{document}

\title{Instant Representation Learning for Recommendation over Large Dynamic Graphs
}


\author{\IEEEauthorblockN{Cheng Wu\IEEEauthorrefmark{1},
Chaokun Wang\IEEEauthorrefmark{1},
Jingcao Xu\IEEEauthorrefmark{1},
Ziwei Fang\IEEEauthorrefmark{1},
Tiankai Gu\IEEEauthorrefmark{1},\\
Changping Wang\IEEEauthorrefmark{2},
Yang Song\IEEEauthorrefmark{2},
Kai Zheng\IEEEauthorrefmark{2},
Xiaowei Wang\IEEEauthorrefmark{2}, and
Guorui Zhou\IEEEauthorrefmark{2}}
\IEEEauthorblockA{\IEEEauthorrefmark{1}School of Software,
Tsinghua University, Beijing 100084, China\\
Email: \{c-wu19, xjc20, fzw20, gtk18\}@mails.tsinghua.edu.cn, chaokun@tsinghua.edu.cn}
\IEEEauthorblockA{\IEEEauthorrefmark{2}Kuaishou Inc., Beijing 100000, China\\
Email: wcpvincent@gmail.com, \{yangsong, zhengkai, wangxiaowei03, zhouguorui\}@kuaishou.com}}

\maketitle
\IEEEpeerreviewmaketitle
\begin{abstract}
Recommender systems are able to learn user preferences based on user and item representations via their historical behaviors.
To improve representation learning, recent recommendation models start leveraging information from various behavior types exhibited by users.
\todo{In real-world scenarios, the user behavioral graph is not only \textit{multiplex} but also \textit{dynamic}, i.e., the graph evolves rapidly over time, with various types of nodes and edges added or deleted, which causes the \textit{Neighborhood Disturbance}.}
Nevertheless, most existing methods neglect such streaming dynamics and thus need to be retrained once the graph has significantly evolved, making them unsuitable in the online learning environment.
\todo{Furthermore, the Neighborhood Disturbance existing in dynamic graphs deteriorates the performance of neighbor-aggregation based graph models.}
To this end, we propose SUPA, a novel graph neural network for dynamic multiplex heterogeneous graphs.
Compared to neighbor-aggregation architecture, SUPA develops a sample-update-propagate architecture to alleviate neighborhood disturbance.
Specifically, for each new edge, SUPA samples an influenced subgraph, updates the representations of the two interactive nodes, and propagates the interaction information to the sampled subgraph.
Furthermore, to train SUPA incrementally online, we propose InsLearn, an efficient \todo{workflow} for single-pass training of large dynamic graphs.
Extensive experimental results on six real-world datasets show that SUPA has a good generalization ability and is superior to \change{sixteen} state-of-the-art baseline methods.
The source code is available at \textit{https://github.com/shatter15/SUPA}.
\end{abstract}


\section{Introduction}
\label{sec:introduction}

\begin{figure}[ht]
\centering
\includegraphics[width=0.8\linewidth]{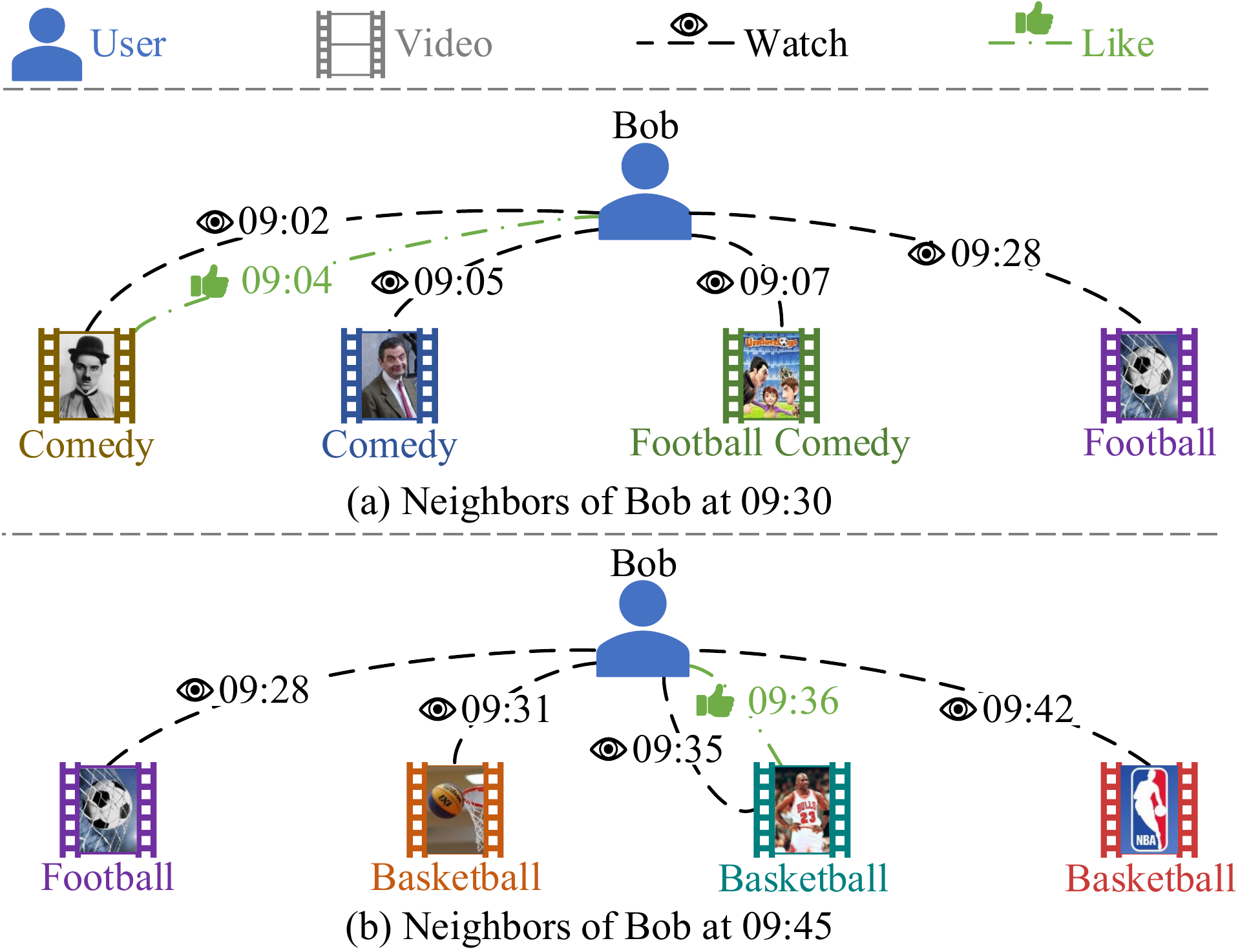}
\caption{A toy example of dynamic multiplex heterogeneous graph for video-watching platforms, where a user can have at most four neighbors due to resource constraints.
It can be observed that the neighbors of Bob at 09:30 and 09:45 share only one common video, which indicates Bob's quick interest drift from comedy to sports, making it difficult for existing static graph-based methods to capture new user interests.
}
\vspace{-15pt}
\label{fig:toy_network}
\end{figure}

\enlargethispage{2em}
Nowadays, recommender systems have become a fundamental service to meet diverse personalized interests of users and alleviate information overload~\cite{ricci2011introduction,li2019routing,jiang2020aspect,li2020quaternion,tan2021multi,zhao2021variational,yin2019social,song2020poisonrec,guo2020group, deng2021knowledge}. As one of the most extensive acceptable paradigms to build a recommendation model, Collaborative Filtering (CF) learns latent representations of users and items based on their historical interactions, thus being able to figure out user preferences through predicting possible links between users and items~\cite{chen2020efficient,he2017neural,xue2017deep,du2018collaborative,sedhain2015autorec}.

Recently, graph modeling has become new state-of-the-art for CF~\cite{wang2019kgat,zhang2019star,he2020lightgcn}. Specifically, graph modeling for recommendation constructs graphs by regarding users, items and auxiliary knowledge graph entities as nodes, and the interactions among them as edges. In this way, graph embedding methods and graph neural networks can be leveraged to model the complex interaction patterns between users and items, and to explore non-linear feature representations for users and items. The learned representations are then used to make recommendations based on the link prediction task.
Early graph modeling for recommendation only \change{considers} a single type of user \change{behaviors} (e.g., purchase behavior on e-commerce platform or click behavior on video-watching platform) to characterize user-item collaborative relations, which neglects the multiplex heterogeneity of the behavioral graphs in recommender systems.
To address this issue, several latest studies pay attention to distilling collaborative signals from the multiplex user-item interactive behaviors, and are able to boost the recommendation performance~\cite{xia2021knowledge,jin2020multi,xia2021graph}.

\enlargethispage{2em}
Nevertheless, existing methods are not optimal in real-world applications where online data are produced in a streaming manner with large scale.
For example, there are over 300 million daily active users on Kuaishou, a short video watching platform in 2021, and each daily active user views short videos for about two hours per day on average\footnote{https://ir.kuaishou.com/news-releases/news-release-details/kuaishou-technology-announces-third-quarter-2021-financial}.
Existing methods fail to deal with data in this scale well, and the reasons are mainly \change{two-fold}.
On one hand, most of the existing methods require to iterate on the training data for multiple epochs to learn node representations, even if they are designed for dynamic graphs~\cite{pareja2020evolvegcn,ma2020streaming,yin2019dhne}.
As a result, they need to be retrained once the graph has a large evolution.
For instance, as shown in Figure~\ref{fig:toy_network}, from 09:30 to 09:45, Bob exhibits an instant interest drift from comedy to sports and the model learned at 09:30 must be retrained or fine-tuned to capture this information.
On the other hand, in real-world scenarios, due to machine memory and storage limitations, only the most recent subgraph is available, \todo{i.e., some outdated nodes and edges are deleted}.
Furthermore, as the graph is updated rapidly, the sampled neighbors of a same node at two moments may have a remarkable difference.
We call this phenomenon as \textit{Neighborhood Disturbance}, and it empirically results in performance degradation for graph models that are based on aggregating information from nodes' neighbors.
For example, in Figure~\ref{fig:toy_network}, the neighbors of Bob at 09:30 and 09:45 share only one common video (i.e., Football) due to machine capability. In this regard, the neighbor-aggregation based models may get two dramatically different embeddings for Bob, which fails to represent Bob's long-term preference.

Thus, it is necessary to explore a more practical representation learning method for large dynamic graphs. However, it is non-trivial,
and there are two major challenges:

\noindent\textbf{Modeling dynamic multiplex heterogeneous graphs.}
In real-world recommender systems, graphs are not only dynamic but also multiplex heterogeneous where multiple types of edges exist between nodes (e.g., Bob watches and likes comedy in Figure~\ref{fig:toy_network}).
In this situation, different types of nodes evolve in different manners, while the impacts of different types of edges also decrease over time. Thus\change{,} it is necessary to distinguish different types of nodes and edges when learning representations.

\enlargethispage{2em}
\noindent\textbf{Dealing with rapid graph evolution.}
Ideally, models are supposed to be fine-tuned instantly after the graph updates.
\todo{Take the short video watching platform, Kuaishou, \change{as an example}. There are millions of videos uploaded each day, and most videos fail to interest users anymore after several hours since they were published. Thus it is necessary to learn and update video representations for recommendation as soon as possible.}
However, most dynamic graph representation learning methods~\cite{pareja2020evolvegcn,ma2020streaming,yin2019dhne} are trained considering the established order of edges, but neglect how to keep the model \change{up-to-date}.

\begin{figure*}[ht]
\centering
\includegraphics[width=0.8\linewidth]{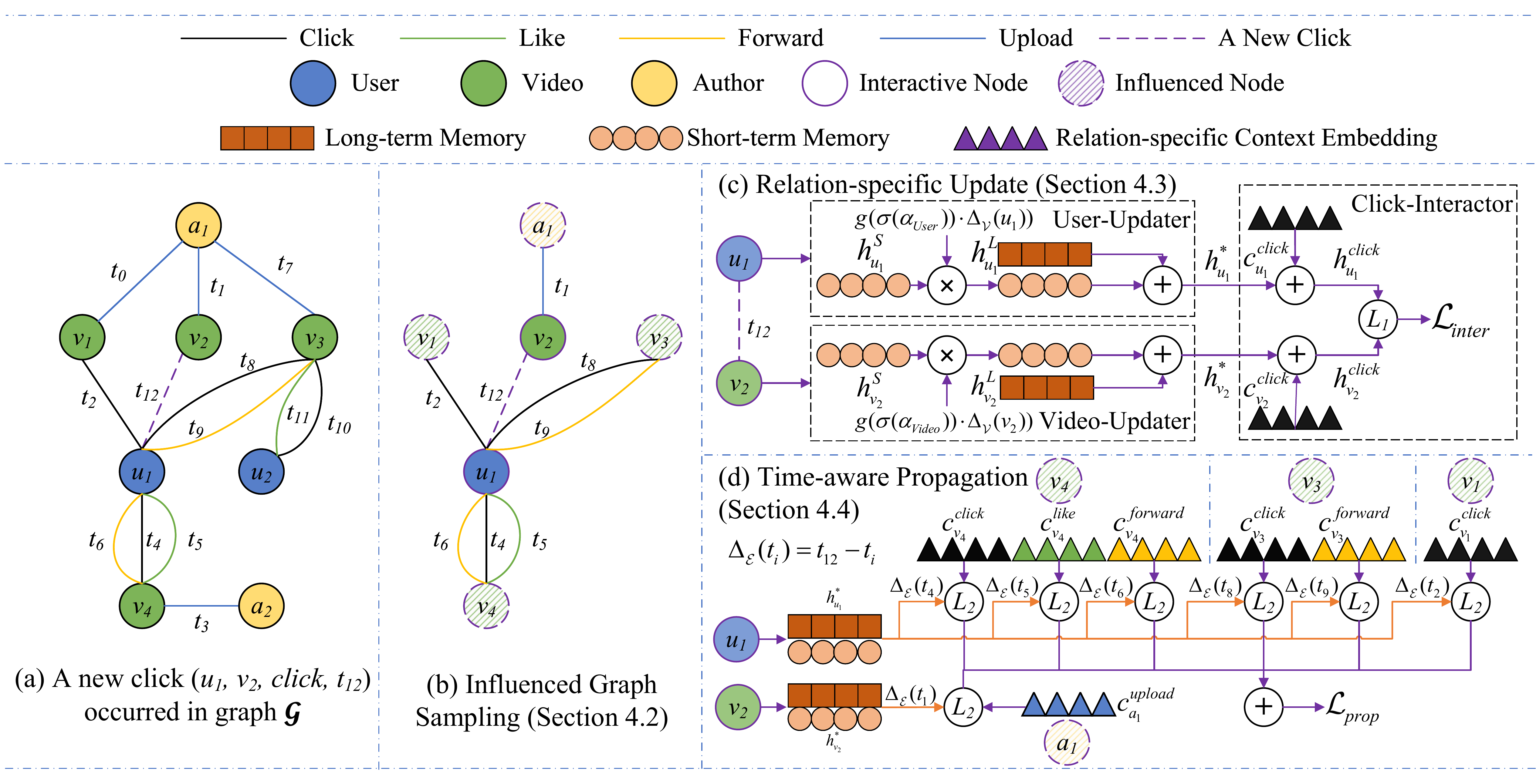}
\caption{The architecture of SUPA (best viewed in color). For a new edge $e = (u_1, v_2, click, t_{12})$, the \textit{Influenced Graph Sampling Module} samples paths
to construct the influenced graph $\mathcal{G}_{s, e}$. Then the \textit{Relation-specific Update Module} updates the representations of the interactive nodes considering the semantic and time information of the edge. After that, the \textit{Time-aware Propagation Module} propagates the interaction information on $\mathcal{G}_{s, e}$ to update the representations of the influenced nodes.}
\label{fig:architure}
\vspace{-10pt}
\end{figure*}

To tackle the above challenges, we propose SUPA, a novel graph neural network (GNN) model for dynamic multiplex heterogeneous graphs (DMHGs) based on a \textbf{S}ample-\textbf{U}pdate-\textbf{P}ropagate \textbf{A}rchitecture (see Figure~\ref{fig:architure}).
Compared to neighbor-aggregation based graph models, SUPA can alleviate neighborhood disturbance since it does not directly aggregate information from neighbors.
To model DMHGs, SUPA develops three techniques:
1) using components with different parameters to handle different types of nodes and edges,
2) splitting node representation into long- and short-term memories, and forgetting short-term memory according to the active time interval when updating,
and 3) decreasing the interaction information according to the interactive time interval while propagating.
To enable SUPA to run online with rapid graph evolution, we propose InsLearn, an \textbf{Ins}tant representation \textbf{Learn}ing \todo{workflow} to train SUPA incrementally within a single pass. Specifically, InsLearn separates the training data into batches according to timestamps and then SUPA is trained and validated on these batches sequentially. In this way, SUPA obtains time-specific and relation-specific node representations to make recommendations.

The main contributions of this paper are highlighted as follows:
\begin{itemize}
    \item We first explore the necessity to learn instant representations over large dynamic graphs for recommendation, and formally define the problem of \textit{DMHG \todo{Instant} Representation Learning} (Section~\ref{sec:definition}).
    \item \todo{We propose the SUPA model to solve the DMHG instant representation learning problem, which alleviates neighborhood disturbance and is able to model the multiplex heterogeneity and streaming dynamics simultaneously.
    We also develop InsLearn, a practical training \todo{workflow} to train SUPA incrementally within a single pass (Section~\ref{sec:methods}).}
    \item We conduct comprehensive experiments to evaluate SUPA with \change{sixteen} state-of-the-art baseline methods.
    The experimental results show the superiority and the generalization ability of SUPA (Section~\ref{sec:experiments}).
\end{itemize}

\begin{table}[t]
\fontsize{7}{7.5}\selectfont
\centering
\caption{\todo{Notations in this paper.}}
\begin{tabular}{c|l}
\toprule
Notations & Definitions \\
\midrule
$\mathcal{O}$ & the set of node types w.r.t. a DMHG \\
$\mathcal{R}$ & the set of edge types w.r.t. a DMHG \\
$|\mathcal{O}|, |\mathcal{R}|$ & the number of node type and edge types, respectively \\
$\mathcal{V}$ & the set of nodes of a DMHG \\
$\mathcal{E}$ & the set of edges of a DMHG, $\mathcal{E}\subseteq
\mathcal{V}\times\mathcal{V}\times\mathcal{R}\times\mathbb{R}^+$\\
$\mathcal{T}$ & the set of timestamps, $\mathcal{T} = \{t \mid (u, v, r, t) \in \mathcal{E}\}$ \\
$\mathcal{G}$ & a DMHG $\mathcal{G}= (\mathcal{V}, \mathcal{E}, \mathcal{O}, \mathcal{R})$ \\
$\phi$ & the node type mapping function of a DMHG, $\phi: \mathcal{V}\to\mathcal{O}$ \\
\midrule
$d$ & the dimension of node representations \\
$\mathcal{P}$ & a multiplex metapath schema \\
\change{$|\mathcal{P}|$} & \change{the length of the multiplex metapath schema} \\
$\vec{\mathbf{\mathcal{P}}}$ & a predefined set of multiplex metapath schemas \\
\change{$p$} & \change{a sampled path} \\ 
$k$ & the number of paths (i.e., walks) to sample for each node  \\
$l$ & the walk length \\
$\mathcal{G}_{s,e}$ & the influenced graph w.r.t. a new edge $e$ \\
$\tau$ & the filter threshold to terminate a propagation flow \\
$N_{neg}$ & the number of negative sampling nodes \\
\midrule
$h^L_v \in \mathbb{R}^{d\times 1}$ & the long-term memories of node $v$ \\
$h^S_v \in \mathbb{R}^{d\times 1}$ & the short-term memories of node $v$ \\
$h^*_v \in \mathbb{R}^{d\times 1}$ & the target embeddings of node $v$ \\
$c^r_v \in \mathbb{R}^{d\times 1}$ & the context embeddings of node $v$ w.r.t. edge type $r$ \\
$h^r_v \in \mathbb{R}^{d\times 1}$ & the final embeddings of node $v$ w.r.t. edge type $r$ \\
$d_{p, v} \in \mathbb{R}^{d\times 1}$ & the interaction information at node $v$ w.r.t. path $p$ \\
\midrule
$N_{iter}$ & the maximum iteration number \\
$I_{valid}$ & the validation interval \\
\change{$\mathcal{E}_{train}$} & \change{the validation set within each batch} \\
\change{$\mathcal{E}_{valid}$} & \change{the training set within each batch} \\
\change{$S_{batch}$} & \change{the size of each training batch} \\
$S_{valid}$ & the size of validation set \\
$\mu$ & the patience for early stopping \\
\bottomrule
\end{tabular}
\label{tab:notations}
\vspace{-5pt}
\end{table}

\enlargethispage{2em}
\section{Definitions}
\label{sec:definition}
In this section, we give some definitions used in this paper.
\begin{definition}[Dynamic Multiplex Heterogeneous Graph]
Let $\mathcal{O}$ and $\mathcal{R}$ denote the sets of node types and edge types, respectively. Then a dynamic multiplex heterogeneous graph (DMHG) can be defined as $\mathcal{G} = (\mathcal{V}, \mathcal{E}, \mathcal{O}, \mathcal{R})$ associated with a node type mapping function $\phi: \mathcal{V} \rightarrow \mathcal{O}$. Here, $\mathcal{V}$ is the set of nodes and $\mathcal{E} \subseteq \mathcal{V} \times \mathcal{V} \times \mathcal{R} \times \mathbb{R}^+$ is the set of temporal edges between nodes in $\mathcal{V}$. At the finest granularity, each edge $(u, v, r, t) \in \mathcal{E}$ may be assigned a unique timestamp $t \in \mathbb{R}^+$.
\end{definition}

Notice that the definition of DMHG is general and can be also used to represent static graphs, homogeneous graphs and non-multiplex heterogeneous graphs. Specifically, for static graphs, all the edges in $\mathcal{E}$ are associated with the same timestamp; for homogeneous graphs, we have $|\mathcal{O}| = |\mathcal{R}| = 1$; for non-multiplex heterogeneous graphs, we have $|\mathcal{O}| + |\mathcal{R}| > 1$ and for each edge $(u, v, r, t) \in \mathcal{E}, r' \in \mathcal{R}, r' \ne r, t' \in \mathbb{R}^+$, we have $(u, v, r', t') \notin \mathcal{E}$.

In recommender systems, $\mathcal{O}$ usually consists of user and item, while $\mathcal{R}$ is the behaviors exhibited by users.
In reality, edges in a DMHG arrive successively and each edge not only provides information of its two nodes, but also has some impact on its local context (i.e., the neighbor nodes and edges). For convenience, we call the two nodes involved in an interaction (i.e., edge) as the \textit{interactive nodes}, while the nodes in the local context are called the \textit{influenced nodes}.

\enlargethispage{2em}
\todo{\begin{definition}[Dynamic Multiplex Heterogeneous Graph Instant Representation Learning Problem]
Given a DMHG $\mathcal{G} = (\mathcal{V}, \mathcal{E}, \mathcal{O}, \mathcal{R})$, the problem of DMHG instant representation learning is to learn a function $\varphi: \mathcal{V} \times \mathcal{R} \times \mathcal{T} \to\mathbb{R}^d$ to map each node $v \in \mathcal{V}$ to a relation-specific $d$-dimensional vector at a given timestamp, where $\mathcal{T} = \{t \mid (u, v, r, t) \in \mathcal{E}\},d \ll|\mathcal{V}|$.
\end{definition}}

\todo{Note that DMHG instant representation learning aims to learn a relation-specific low-dimensional vector for each node at any timestamp where there is at least one edge established, i.e., models solving the DMHG instant representation learning problem are expected to update node representations if necessary once there is a new edge occurring.}


\begin{figure}[ht]
\centering
\includegraphics[width=0.75\linewidth]{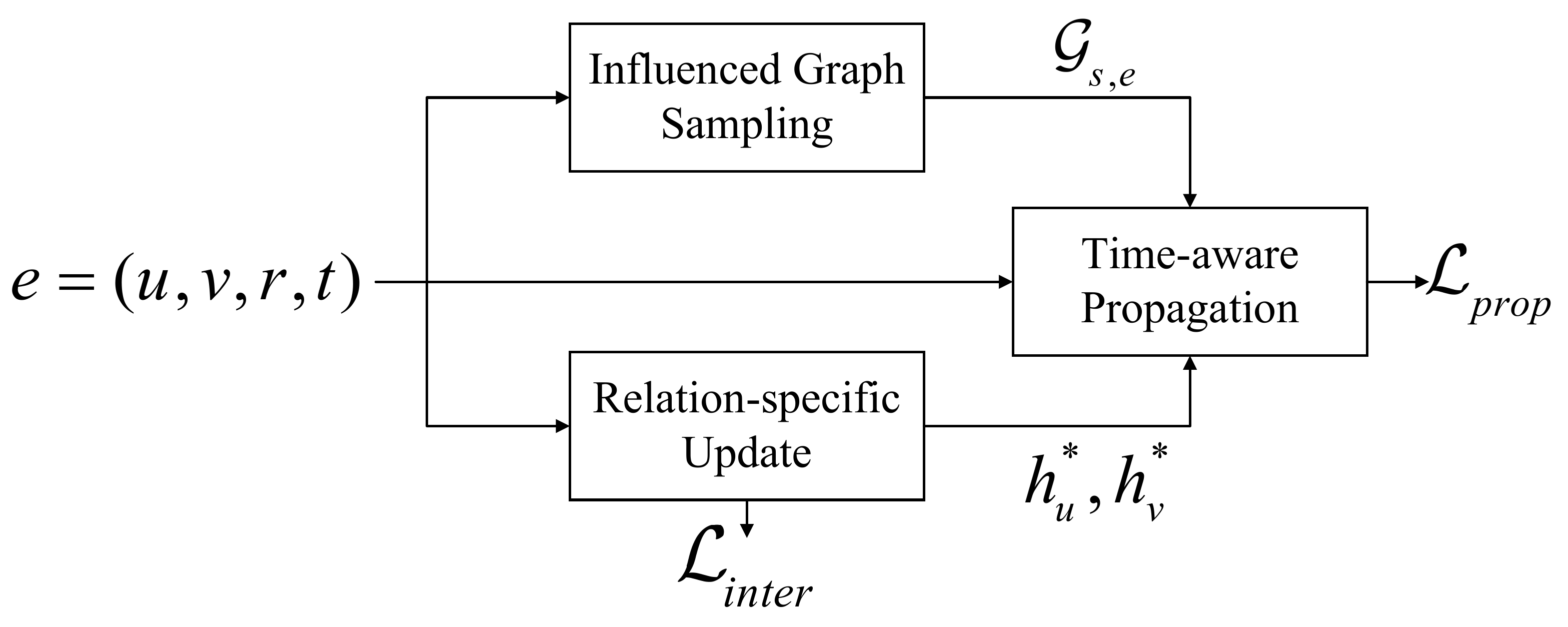}
\vspace{-5pt}
\caption{\change{The interactions among different modules during training.}}
\label{fig:interactions}
\vspace{-10pt}
\end{figure}

\begin{definition}[Multiplex Metapath Schema]
A multiplex metapath schema $\mathcal{P}$ is defined as a path with specified node types and edge types, denoted as $\mathcal{P} = o_1\stackrel{R_1}{\rightarrow}o_2\stackrel{R_2}{\rightarrow} \cdots \stackrel{R_{n - 1}}{\rightarrow}o_n$. Here, $o_i \in\mathcal{O}~(i = 1,2,\cdots,n)$ is the node type, $R_j \subseteq \mathcal{R}~(j = 1, 2, \cdots, n - 1)$ is the set of edge types, and $|\mathcal{P}|=n$ is the length of the multiplex metapath schema.
\end{definition}

\todo{The multiplex metapath schema is capable to describe the multi-hop and high-order relationships between nodes in multiplex heterogeneous networks. For example, in the network shown in Figure~\ref{fig:toy_network}, the multiplex metapath schema $User\stackrel{\{click, like\}}{\longrightarrow}Video\stackrel{\{upload\}}{\longrightarrow}Author\stackrel{\{upload\}}{\longrightarrow}Video\stackrel{\{click, like\}}{\longrightarrow}User$ represents the user-user relationship in which users have watched or liked videos published by the same author. Note that in this paper, the multiplex metapath schemas are predefined according to the characteristics of datasets,} \review{and we leave the automatic mining of metapath schemas in future work.}

\todo{Several important notations used in this paper are summarized in Table~\ref{tab:notations}.}

\vspace{-5pt}
\enlargethispage{2em}
\section{Methods}
\label{sec:methods}
In this section, we first present the overview of the proposed model SUPA, and bring the details of its three modules. Then, we propose the training procedure of SUPA.
\change{At last, we discuss the application of our method, analyze the training time complexity and make a comparison with existing work.}

\begin{table}[ht]
    \scriptsize
    \centering
    \caption{\change{An example of influenced graph sampling.}}
    \vspace{-5pt}
    \begin{tabular}{|cccccc|}
    \hline
    \multicolumn{6}{|c|}{$\mathcal{P} = User\stackrel{\{click\}}{\longrightarrow}Video\stackrel{\{click\}}{\longrightarrow}User$} \\
    \hline
    \multicolumn{6}{|c|}{ $p = u_2\stackrel{click}{\longrightarrow}v_3\stackrel{click}{\longrightarrow}u_1\stackrel{click}{\longrightarrow}v_1\stackrel{click}{\longrightarrow}u_1$} \\
    \hline
    \hline
    $i$ & 1 & 2 & 3 & 4 & 5 \\
    \hline
    $f(i, |\mathcal{P}| - 1)$ & 1 & 2 & 1 & 2 & 1 \\
    $o_{\mathcal{P}, f(i, |\mathcal{P}| - 1})$ & $User$ & $Video$ & $User$ & $Video$ & $User$ \\
    $\phi(p_i)$ & $User$ & $Video$ & $User$ & $Video$ & $User$ \\
    \hline
    \hline
    $j$ & 1 & 2 & 3 & 4 & \\
    \hline
    $f(j, |\mathcal{P}| - 1)$ & 1 & 2 & 1 & 2 & \\
    $R_{\mathcal{P}, f(j, |\mathcal{P}| - 1)}$ & $\{click\}$ & $\{click\}$ & $\{click\}$ & $\{click\}$ & \\
    $\psi(p_{j}, p_{j + 1})$ & $click$ & $click$ & $click$ & $click$ & \\
    \hline
    \end{tabular}
    \label{tab:influenced}
    \vspace{-15pt}
\end{table}

\subsection{Model Overview}
As shown in Figure~\ref{fig:architure} and Figure~\ref{fig:interactions}, the architecture of SUPA consists of three modules: the \textit{Influenced Graph Sampling Module}, the \textit{Relation-specific Update Module} and the \textit{Time-aware Propagation Module}.

For a new edge, the \textit{Influenced Graph Sampling Module} samples paths starting from the two interactive nodes respectively based on a given multiplex metapath schema set to construct an influenced graph, which is further used for later processing.
In the \textit{Relation-specific Update Module}, the node-type specific updater first updates the embeddings of the two interactive nodes by forgetting the corresponding short-term memory according to the active time intervals, while the edge-type specific interactor calculates the interaction loss w.r.t. the edge type.
The interaction information is then propagated in the \textit{Time-aware Propagation Module}, which decreases the interaction information according to the interactive time intervals. The propagation loss is also computed in this module.
Combining the interaction loss and the propagation loss, SUPA is trained incrementally, and can give different most-recent node embeddings w.r.t. different interaction (i.e., edge) types.

Note that although the modules are designed for DMHGs, SUPA is able to deal with static networks or homogeneous networks, too. Specifically, for static networks, SUPA can be trained by setting a same timestamp for all the interactions, while for homogeneous networks, we just need to set the number of node types and edge types to one, respectively.

\todo{Also, SUPA can deal with edge deletion although it is designed to consider the case of adding new edges.
The reasons are \change{two-fold}.
On one hand, the time-aware propagation module takes the time interval into consideration and will not propagate information through outdated (i.e., deleted) edges.
On the other hand, edge deletion can be viewed as a special relation (i.e., edge type) among nodes, and thus shares the same process procedure with edge addition.}


\enlargethispage{2em}
\subsection{\todo{Influenced Graph Sampling}}
\label{sec:active_graph_sampling}
The \textit{Influenced Graph Sampling Module} is designed to determine the influenced nodes for each interaction.
For example, Figure~\ref{fig:architure}b shows a sampled influenced graph of Figure~\ref{fig:architure}a w.r.t a new edge $e = (u_1, v_2, click, t_{12})$.
Formally, for a new edge $(u, v, r, t)$, given a predefined multiplex metapath schema set $\vec{\mathbf{\mathcal{P}}} = \{\mathcal{P}_1, \mathcal{P}_2, \ldots, \mathcal{P}_s\}$, we sample two path sets starting from $u$ and $v$, respectively:
\begin{small}
\vspace{-2pt}
\begin{align}
    \vec{\mathbf{p}}_u = \{\mathbf{p}_u^1, \mathbf{p}_u^2, \cdots, \mathbf{p}_u^k\},
    \ \ \ \
    \vec{\mathbf{p}}_v = \{\mathbf{p}_v^1,, \mathbf{p}_v^2, \cdots, \mathbf{p}_v^k\},
\end{align}
\end{small}where $k$ is the number of paths. For each $p \in (\vec{\mathbf{p}}_u \cup \vec{\mathbf{p}}_v)$, there is a multiplex metapath schema $\mathcal{P} \in \vec{\mathbf{\mathcal{P}}}$ satisfying that:
\begin{small}
\vspace{-2pt}
\begin{align}
    \phi(p_i) &= o_{\mathcal{P}, f(i, |\mathcal{P}| - 1)},& i = 1, 2, \cdots, l, \\
    \psi(p_{j}, p_{j + 1}) &\in R_{\mathcal{P}, f(j, |\mathcal{P}| - 1)},& j = 1, 2, \cdots, l - 1,
\end{align}
\end{small}where $|p| = l$, $l$ is the predefined walk length, $p_i$ is the $i$-th node in $p$, $\phi(u)$ is the node type of $u$, $\psi(u, v)$ is the edge type between nodes $u$ and $v$, $f(i, L) = ((i - 1) \bmod L) + 1$, $o_{\mathcal{P}, i}$ is the $i$-th node type and $R_{\mathcal{P}, i}$ is the $i$-th set of edge types in $\mathcal{P}$.

\enlargethispage{2em}
\change{Table~\ref{tab:influenced} shows an example of influenced graph sampling w.r.t. Figure~\ref{fig:architure}a in which $\mathcal{P} = User\stackrel{\{click\}}{\longrightarrow}Video\stackrel{\{click\}}{\longrightarrow}User$ and a path starting from $u_2$ satisfying $\mathcal{P}$ with $l = 5$ is $p = u_2\stackrel{click}{\longrightarrow}v_3\stackrel{click}{\longrightarrow}u_1\stackrel{click}{\longrightarrow}v_1\stackrel{click}{\longrightarrow}u_1$.}
Notice that the walk length $l$ is not necessary to be equal to the length of the multiplex metapath schema since we repeat the multiplex metapath schema by treating the tail node type as the head node type, which is \change{the reason for using modulus $|\mathcal{P}| - 1$ instead of $|\mathcal{P}|$} and is meaningful only for symmetrical schema. However, we can easily convert an asymmetrical schema $\mathcal{P} = o_1\stackrel{R_1}{\rightarrow}\cdots\stackrel{R_{n-2}}{\rightarrow}o_{n-1}\stackrel{R_{n - 1}}{\rightarrow}o_n$ to a symmetrical one:
\begin{small}
\begin{align}
\mathcal{P}' = o_1\stackrel{R_1}{\rightarrow}\cdots\stackrel{R_{n-2}}{\rightarrow}o_{n-1}\stackrel{R_{n - 1}}{\rightarrow}o_n \stackrel{R_{n - 1}}{\rightarrow}
o_{n-1}\stackrel{R_{n-2}}{\rightarrow}\cdots\stackrel{R_1}{\rightarrow}o_1.
\end{align}
\end{small}

Collecting all the sampling paths w.r.t. a new edge $e$ together, we obtain the influenced graph $\mathcal{G}_{s, e}$, which is used in Section~\ref{sec:time_aware_propagation}.
\subsection{Relation-specific Update}
\label{sec:relation_specific_update}
In this subsection, we introduce the two components of the \textit{Relation-specific Update Module}, which are designed to update the two interactive nodes for each edge considering the multiplex heterogeneity. As shown in Figure~\ref{fig:architure}c, for the new edge $(u_1, v_2, click, t_{12})$, the node-type specific updater updates the \textit{short-term memories} of nodes, while the edge-type specific interactor combines the embeddings of the two interactive nodes to calculate the interaction loss. More precisely, for each node in the graph, we maintain three kinds of learnable vectors, namely \textit{the long-term memory}, \textit{the short-term memory} and \textit{the relation-specific context embedding}, which are initialized randomly. The updater utilizes a monotone decreasing function to forget the short-term memory according to the active time interval, while keeping the long-term memory unchanged, which aims to model the drift of node features. The interactor then combines the memories of nodes with the corresponding relation-specific context embedding to calculate the interaction loss.

\subsubsection{Node-type Specific Updater}
For a new edge $(u, v, r, t)$, given the long-term memories $h_u^L, h_v^L$ and short-term memories $h_u^S, h_v^S$ for node $u, v$, respectively, the updater calculates the target embeddings $h_u^*, h_v^*$ as:
\begin{small}
\begin{align}
    h_u^* &= h_u^L + h_u^S\cdot g(\sigma(\alpha_{\phi(u)}) \cdot \Delta_\mathcal{V}(u)), \notag\\
    h_v^* &= h_v^L + h_v^S\cdot g(\sigma(\alpha_{\phi(v)}) \cdot \Delta_\mathcal{V}(v)),
\end{align}
\end{small}where $g(\cdot)$ is a decreasing function defined as $g(x) = \frac{1}{\log(e + x)}$, ($\alpha_o$)s ($o \in \mathcal{O}$, \change{$\alpha_o \in \mathbb{R}$}) are the node-type specific parameters to learn,
\change{$\sigma(x) = \frac{1}{1 + e^{-x}}$ is the sigmoid function to limit the coefficient of $\Delta_\mathcal{V}(i)$ (i.e., $\sigma(\alpha_o)$) to $(0, 1)$, }
$\Delta_\mathcal{V}(i) = t - t'_i$ \change{($\Delta_\mathcal{V}(i) > 0$)} is the active time interval of node $i$, and $t'_i$ is the timestamp of the latest interaction involving node $i$.

Notice that we set different ($\alpha_o$)s for different node types because the memory drift for nodes of different kinds might distinguish from each other. For example, in an e-commerce network, the short-term memory of a user is more likely to change compared with the one of an item.

\subsubsection{Edge-type Specific Interactor}
We utilize the relation-specific context embedding to model the inspiration that different kinds of interaction introduce diverse information of node features. Specifically, for a given node $v \in \mathcal{V}$ and an edge type $r \in \mathcal{R}$, the context embedding of $v$ w.r.t. $r$ is $c_v^r$. Combing the target embedding (i.e., the output of the updater) and the context embedding, we calculate the final embeddings of $u$ and $v$ as follows:
\begin{small}
\begin{align}
    h_u^r = \frac{1}{2}(h_u^* + c_u^r),
    \ \ \ \
    h_v^r = \frac{1}{2}(h_v^* + c_v^r).
\end{align}
\end{small}

Note that via calculating $h_u^r$ and $h_v^r$ according to the edge type, the interactor can capture the multiplex heterogeneity.
The interaction loss of edge $(u, v, r, t)$ is defined as follows:
\begin{small}
\begin{align}
    \mathcal{L}_{inter} = -\log{\sigma({h_u^r}^\top h_v^r)},
\end{align}
\end{small}where $\sigma(\cdot)$ is the sigmoid function. Intuitively, the interaction loss aims to make the final embeddings of the two interactive nodes as similar to each other as possible.

\enlargethispage{2em}
\subsection{Time-aware Propagation}
\label{sec:time_aware_propagation}


In this subsection, we present the propagation procedure with two time-specific considerations, namely 1) \textbf{attenuation}, where the information is decreased when propagating across old edges, and 2) \textbf{termination}, where the propagation flow is stopped when encountering an out-of-date edge.
Intuitively, there are three steps in the procedure: initialization,  propagation and loss computation.
As shown in Figure~\ref{fig:architure}d, there are two propagation flows originating from the two interactive nodes (i.e., $u_1$ and $v_2$), respectively.
Specifically, using the target embedding of $u_1$ as the initial interaction information, the first propagation flow starts from $u_1$ and propagates to the influenced node in $\vec{\mathbf{p}_{u_1}}$ (i.e., $v_1, v_3, v_4$).
Similarly, the second propagation flow starts from $v_2$ and propagates the target embedding of $v_2$ to $a_1$.
After the propagation flows are over, the propagation loss is calculated.

\textbf{Initialization.}
Formally, for a new edge $e = (u, v, r, t)$, the initial interaction information for the propagation flows starting from node $u$ and $v$ is the target embedding $h_u^*$ and $h_v^*$, respectively, since the interaction information is already encoded into the short-term memories of node $u$ and $v$.
Then, the interaction information is propagated on the influenced graph $\mathcal{G}_{s, e}$ (i.e., $\vec{\mathbf{p}_u}$ for the propagation flow starting from $u$ and $\vec{\mathbf{p}_v}$ for the propagation flow starting from $v$).
Note that we do not propagate the information to the whole network for two considerations.
One is that the impact of an edge in a streaming graph is often local~\cite{chang2017streaming}, while the other is that the information propagated to an influenced node $i$ in $\mathcal{G}_{s, e}$ can be further propagated once node $i$ interacts with other nodes.

\textbf{Propagation}.
The interaction information undergoes a slow attenuation related to the interactive time interval when walking through each edge during the propagation flow. More precisely, for an edge $e' = (u', v', r', t')$ w.r.t. a path $p \in \mathcal{G}_{s, e}$, suppose that the interaction information arriving at node $u'$ is $d_{p,u'}$, then the interaction information propagated to $v'$ is:
\begin{small}
\begin{align}
    d_{p, v'} = \mathcal{D}(\Delta_\mathcal{E}(t')) \cdot g(\Delta_\mathcal{E}(t')) \cdot d_{p,u'},
\end{align}
\end{small}where $\Delta_\mathcal{E}(t') = t - t'$ is the time interval from $e_p$ happening to now. $g(\cdot)$ is the same decreasing function as defined in the node-type specific updater (see Section~\ref{sec:relation_specific_update}) to implement the \textbf{attenuation}.
In the process of propagating through edges, the \textbf{termination} is taken into considerations as well. Intuitively, propagating information through extremely early interactive edges might introduce noises. For example, it is unwise to recommend napkins to a man who has bought them ten years ago. Therefore, we use a filter function $\mathcal{D}(\cdot)$ to detect out-of-date edges and terminate such propagation flow:
\begin{small}
{\setlength\abovedisplayskip{2pt}
\setlength\belowdisplayskip{2pt}
\begin{align}
    \mathcal{D}(x) = \left\{\begin{aligned}
        &1, &x \le \tau, \\
        &0, &otherwise,
    \end{aligned}
    \right.
\end{align}}
\end{small}where $\tau$ is a predefined threshold. Briefly, if the interactive time interval is larger than $\tau$, \todo{edge $e'$ is supposed to be outdated (i.e., deleted)}, and the interaction information is set to $\vec{\mathbf{0}}$. Thus, the propagation flow of that path is terminated.

\begin{algorithm}[t]
\scriptsize
\caption{The InsLearn training workflow.}
\label{alg:training_framework}
\LinesNumbered
\KwIn{
The untrained model $\Phi$; The edge set $\mathcal{E}$; The batch size $S_{batch}$; The maximum iteration number $N_{iter}$; The validation interval $I_{valid}$; The size of validation set $S_{valid}$; The patience for early stopping $\mu$
}
\KwOut{The trained model $\Phi$}
Sort $\mathcal{E}$ according to the establishing timestamps of edges\;
$\vec{\mathbf{\mathcal{E}}}_{batch} \leftarrow generate\_sequential\_batches(\mathcal{E}, S_{batch})$\;
\For{$\mathcal{E}_{batch}$ in $\vec{\mathbf{\mathcal{E}}}_{batch}$}{
 Let best score $\theta_{best} \leftarrow 0$, current patience $\mu_{cur} \leftarrow 0$, best model $\Phi_{best} \leftarrow \Phi$ \;
 Let training and validation set $\mathcal{E}_{train}, \mathcal{E}_{valid}\leftarrow split\_train\_valid\_edges(\mathcal{E}_{batch}, S_{valid})$\;
 \For{$i = 1\cdots N_{iter}$}{
    $\Phi.train(\mathcal{E}_{train})$ \;
    \If{$i \bmod I_{valid} = 0$}{
        Let score $\theta \leftarrow \Phi.valid(\mathcal{E}_{valid})$ \;
        \eIf{$\theta > \theta_{best}$}{
            $\theta_{best} \leftarrow \theta$,
            $\Phi_{best} \leftarrow \Phi$,
            $\mu_{cur} \leftarrow 0$ \;
        }{
            $\mu_{cur} \leftarrow \mu_{cur} + 1$ \;
            \If{$\mu_{cur} > \mu$}{break \;}
        }
    }
 }
 $\Phi = \Phi_{best}$\;
}
\Return{$\Phi$}
\end{algorithm}
\setlength{\textfloatsep}{0pt}

\enlargethispage{2em}
\textbf{Loss computation.}
The information propagated across $\mathcal{G}_{s, e}$ is used to calculate the propagation loss \change{$\mathcal{L}_{prop}$}, which is designed to update the representations of the influenced nodes.
\change{Inspired by the skip-gram model~\cite{mikolov2013distributed}, which maximizes the probability of the co-occurrence of a word and its context, the objective of $\mathcal{L}_{prop}$ is to maximize the probability of the co-occurrence of an edge $e = (u,v ,r, t)$ and its influenced graph $\mathcal{G}_{s, e}$ through maximizing the similarity between the context embedding of the influenced nodes and the interaction information, i.e., to minimize the following loss:}
\begin{small}
\begin{align}
    \mathcal{L}_{prop} = -
    \sum_{p \in (\vec{\mathbf{p}}_u \cup \vec{\mathbf{p}}_v)}\sum_{<z_i, r_i>\in p}(\mathcal{X}(d_{p, z_i} \ne \vec{\mathbf{0}}) \cdot 
    \log{\sigma({c_{z_i}^{r_i}}^\top d_{p, z_i})}),
\end{align}
\end{small}where $<z_i, r_i>$ refers to the propagation step to $z_i$ through an edge with type $r_i$, $\mathcal{X}(\cdot)$ is the indicative function, $d_{p, z_i}$ is the interaction information arrived at node $z_i$ w.r.t. path $p$, $c_{z_i}^{r_i}$ is the context embedding of node $z_i$ w.r.t. edge type $r_i$, and $\sigma(\cdot)$ is the sigmoid function.

\subsection{Model Training}
\label{subsec:model_learning}
\todo{In this subsection, we first present the parameter optimization procedure of SUPA, and then propose the InsLearn training \todo{workflow} to train SUPA in a single pass.}
\subsubsection{Parameter Optimization}
For each new edge $(u, v, r, t)$, the objective is to minimize the following loss:
\begin{small}
\begin{align}
\mathcal{L}' = \mathcal{L}_{inter} + \mathcal{L}_{prop}.
\end{align}
\end{small}

Inspired by the negative sampling strategy in the skip-gram model~\cite{mikolov2013distributed}, we design the following negative sampling loss to achieve efficient optimization:
\begin{small}
\begin{align}
\mathcal{L}_{neg} = -&(\sum_{j = 1}^{N_{neg}} E_{k \sim P_{Neg}}[\log{\sigma(-{c_i^r}^\top h_u^*)}] + \notag\\ &\sum_{j = 1}^{N_{neg}} E_{k \sim P_{Neg}}[\log{\sigma(-{c_i^r}^\top h_v^*)}]),
\end{align}
\end{small}where $N_{neg}$ is the number of negative sampling nodes, $P_{Neg}$ is the noise distribution for negative sampling, $i$ is the node sampled from $P_{Neg}$, $\sigma(\cdot)$ is the sigmoid function and $c_i^r$ is the context embedding for node $i$ w.r.t. edge type $r$. Note that we limit $r$ to the same type of the new edge so as to distinguish different context embeddings under different edge types.

Finally, the objective function is revised as:
\begin{small}
\begin{align}
    \mathcal{L} = \mathcal{L}_{inter} + \mathcal{L}_{prop} + \mathcal{L}_{neg}.
\end{align}
\end{small}

We use the back-propagation algorithm to optimize the model parameters and the node representations. After the training procedure is done, we obtain the final embedding of a node $v$ w.r.t. the edge type $r$ as follows:
\begin{small}
\begin{align}
    h_v^r = \frac{1}{2}(h_v^* + c_v^r)  
    = \frac{1}{2}(h_v^L + h_v^S + c_v^r).
\end{align}
\end{small}\change{Intuitively, $h_v^r$ takes the long-term memory, the short-term memory and the relation-specific context embedding into consideration, and is able to capture the time information and the heterogeneous information of node $v$.}

\enlargethispage{2em}
\subsubsection{\todo{Training Workflow}}
As shown in Algorithm~\ref{alg:training_framework}, there are five steps in the InsLearn training \todo{workflow}.
\begin{enumerate}[STEP 1]
    \item Sort the training edges according to their established order, and then divide them into batches with size $S_{batch}$ sequentially (Lines 1--2).
    \item For each batch, take the last $S_{valid}$ edges as the validation set $\mathcal{E}_{valid}$ and the rest edges as the training set $\mathcal{E}_{train}$ (Lines 3--5).
    \item Train the model on $\mathcal{E}_{train}$, and validate the model on $\mathcal{E}_{valid}$ every $I_{valid}$ iterations within each batch. (Lines 6--9).
    \item For each batch, repeat STEP 3 until the performance of the model 1) keeps improving but the number of iterations reaches the predefined maximum value $N_{iter}$, or 2) gets worse continuously and exceeds the patience for early stopping $\mu$ (Lines 10--17).
    \item Select the model with the best validation result to train the next batch or return (Lines 20--22).
\end{enumerate}

There are three considerations in InsLearn.
First, to avoid the problem of over-fitting,
$N_{iter}$ is set to limit the iterations. Once the model is trained on a batch for $N_{iter}$ times, the training procedure will be stopped immediately to alleviate over-fitting even though \change{the performance of the model} still improves.
Second, the early stopping mechanism is adopted to prevent over-training when the model gets well-trained before consuming up the allowed training steps. Generally, over training will introduce extra time cost, while terminating at once when the performance gets worst might result in sub-optimization. To this end, the early stopping mechanism can make a good balance.
Third, for time efficiency, $I_{valid}$ is set to validate the model every several iterations because validating in all iterations will introduce a large number of calculations.

In this way, the model is trained within each batch sequentially. Thus it can be run and updated directly on the real-world platform with streaming data.

\enlargethispage{2em}
\subsection{Model Discussion}
In this subsection, we first present how to apply our proposed method to real-world recommendation, and then analyze the training time complexity of our method. \change{Finally, we make a comparison with existing work.}
\subsubsection{\todo{Making Recommendations}}
The learned node embeddings can be used to make recommendations by predicting possible links between users and items. Specifically, given a user node $u$ and an item node $v$, the possibility that $u$ will interact with $v$ under relation $r$ is as follows:
\begin{small}
\begin{align}
    \gamma(u, v, r) = {h_u^r}^\top h_v^r.
\end{align}
\end{small}

In this way, for all items $v' \in \mathcal{V}, \phi(v') = Item$, we have $\gamma(u, v', r)$, and the set of items that have the top-K $\gamma(u, v', r)$ values are the top-K recommendation results given by SUPA.

\subsubsection{\todo{Training Time Complexity Analysis}}
For each edge, the \textit{Influenced Graph Sampling Module} samples $2k$ paths with length $l$ in total. As each node is sampled with a random choice, the time complexity is $O(kl)$. Then in the \textit{Relation-specific Update Module}, the time complexity to update the embeddings of interactive nodes and calculate $\mathcal{L}_{inter}$ is $O(1)$. After that, the \textit{Time-aware Propagation Module} calculates $\mathcal{L}_{prop}$ according to the sampled paths, and thus the time complexity is $O(kl)$. Besides, \change{$2N_{neg}$} nodes are sampled to compute $\mathcal{L}_{neg}$, which results in an $O(N_{neg})$ time complexity. Finally, since the model is trained by a single iteration on the edge set, the overall time complexity of training SUPA is $O((kl + N_{neg})\cdot|\mathcal{E}|)$.

\subsubsection{\change{Comparison with Existing Work}}
\change{We discuss the differences between our model and existing work, including Graph Recurrent Neural Network (\textbf{GRNN})~\cite{zhou2020graph,seo2018structured,pareja2020evolvegcn} and \textbf{MeLU}~\cite{lee2019melu}.}

\change{\textbf{GRNN}. GRNNs incorporate the recurrent units such as LSTM~\cite{peng2017cross} and GRU~\cite{li2015gated} into GNNs to model time information. GCRN~\cite{seo2018structured} and EvolveGCN~\cite{pareja2020evolvegcn} then extend GRNNs with graph convolutional network and achieve state-of-the-art performances in dynamic networks. With the help of LSTM and GRU, GRNNs are able to capture and select how much knowledge to keep. However, they suffer from the low efficiency of RNN and take a long time to train and evaluate. In this work, we keep the idea of long-term and short-term memories, and remove the recurrent unit for the consideration of efficiency. Besides, we take the heterogeneity into consideration and distinguish different node types when updating node memories.}

\change{\textbf{MeLU}. MeLU~\cite{lee2019melu} is a meta-learning based recommender system which alleviates the cold-start problem. From meta-learning, MeLU can estimate a new user’s preferences based on only a few user-item interactions. What SUPA and MeLU have in common is that they can learn preferences of new users in a few steps. However, MeLU focuses on searching for a group of parameters that are suitable for different users and thus a few updates are enough to find the optimal parameters for a specific user, while SUPA follows the idea of collaborative filtering and develops a novel method to explore the similarities among users and items comprehensively. Furthermore, SUPA takes the time information into consideration and can model the interest drift of existing users, while MeLU ignores such information.}

\enlargethispage{2em}
\section{Experiments}
\label{sec:experiments}
In this section, \todo{the statistics of the datasets, baselines as well as the experimental settings are firstly introduced.}
Then, we evaluate the recommendation performance of SUPA against \change{sixteen} baseline methods with the link prediction task.
Next, we design two experiments to validate the practicability of SUPA, namely dynamic link prediction and link prediction with neighborhood disturbance.
We also conduct ablation studies to show the effect of our proposed components, and \change{investigate the scalability of SUPA in terms of fast changing data.}
\change{Furthermore}, we present the parameter sensitivity analysis.
Finally, we make a visualization to qualitatively illustrate the superiority of SUPA over \change{baseline methods}.

\subsection{Datasets}

\begin{table}
\scriptsize
\centering
\caption{\todo{The statistics of datasets.}}
\vspace{-5pt}
\begin{tabular}{cccccc}
\toprule
Datasets & $|\mathcal{V}|$ & $|\mathcal{E}|$ & $|\mathcal{O}|$ & $|\mathcal{R}|$ & $|\mathcal{T}|$ \\
\midrule
UCI & 1,677 & 56,617 & 1 & 1 & 47,123 \\
Amazon & 10,099 & 148,659 & 1 & 2 & 1 \\
Last.fm & 127,786 & 720,537 & 2 & 1 & 707,959 \\
MovieLens & 16,578 & 1,231,508 & 2 & 2 & 877,684 \\
Taobao & 12,611 & 20,890 & 2 & 4 & 20406 \\
Kuaishou & 138,812 & 1,779,639 & 3 & 5 & 705,302 \\
\bottomrule
\end{tabular}
\label{tab:datasets}
\vspace{-10pt}
\end{table}

\begin{table}
\scriptsize
\centering
\caption{\change{The selected multiplex metapath schemas of datasets.}}
\vspace{-10pt}
\begin{tabular}{c|c}
\toprule
Datasets & $\vec{\mathbf{\mathcal{P}}}$ \\
\midrule
UCI & $U\stackrel{\{C\}}{\longrightarrow}U$ \\
\midrule
Amazon & $P\stackrel{\{L\}}{\longrightarrow}P$ \\
\midrule
Last.fm & $U\stackrel{\{L\}}{\longrightarrow}A\stackrel{\{L\}}{\longrightarrow}U, A\stackrel{\{L\}}{\longrightarrow}U\stackrel{\{L\}}{\longrightarrow}A$ \\
\midrule
MovieLens & $U\stackrel{\{R,T\}}{\longrightarrow}M\stackrel{\{R,T\}}{\longrightarrow}U, M\stackrel{\{R,T\}}{\longrightarrow}U\stackrel{\{R,T\}}{\longrightarrow}M$ \\
\midrule
Taobao & $U\stackrel{\{P,B,C,F\}}{\longrightarrow}I\stackrel{\{P,B,C,F\}}{\longrightarrow}U, I\stackrel{\{P,B,C,F\}}{\longrightarrow}U\stackrel{\{P,B,C,F\}}{\longrightarrow}I$ \\
\midrule
\multirow{2}{*}{Kuaishou} & $U\stackrel{\{W,L,F,C\}}{\longrightarrow}V\stackrel{\{W,L,F,C\}}{\longrightarrow}U, A\stackrel{\{U\}}{\longrightarrow}V\stackrel{\{U\}}{\longrightarrow}A$ \\
& $V\stackrel{\{W,L,F,C\}}{\longrightarrow}U\stackrel{\{W,L,F,C\}}{\longrightarrow}V, V\stackrel{\{U\}}{\longrightarrow}A\stackrel{\{U\}}{\longrightarrow}V$ \\
\bottomrule
\end{tabular}
\label{tab:metapaths}
\end{table}

\begin{table*}
\scriptsize
\caption{The experimental results w.r.t. \textbf{H@K} of link prediction, where ``-'' means that the method fails to give results in a week. The best results are illustrated in bold and the number underlined is the runner-up. Please note that the number with a star ($*$) indicates
\change{the result achieves statistically improvement}
with $p < 0.01$ under t-test compared to other baselines.}
\vspace{-9pt}
\centering
\begin{tabular}{c|cc|cc|cc|cc|cc|cc}
\toprule
& \multicolumn{2}{c|}{UCI} & \multicolumn{2}{c|}{Amazon} & \multicolumn{2}{c|}{Last.fm} & \multicolumn{2}{c|}{MovieLens} & \multicolumn{2}{c|}{Taobao} & \multicolumn{2}{c}{Kuaishou}\\
  & H@20 & H@50 & H@20 & H@50 & H@20 & H@50 & H@20 & H@50 & H@20 & H@50 & H@20 & H@50 \\
\midrule
DeepWalk~\cite{perozzi2014deepwalk} & \change{0.1859} & \change{0.2550} & \change{0.2826} & \change{0.4496} & \change{0.0701} & \change{0.1344} & 0.0276 & 0.0674 & \underline{0.3172} & 0.3522 & 0.0210 & 0.0486  \\
LINE~\cite{tang2015line} & 0.0697 & 0.1086 & \change{0.2210} & \change{0.3327} & \change{0.0455} & \change{0.0857} & \change{0.0321} & \change{0.0584} & \change{0.2850} & \change{0.3117} & \change{0.0106} & \change{0.0223}  \\
node2vec~\cite{node2vec-kdd2016} & 0.1214 & 0.1808 & 0.1970 & 0.3026 & \change{0.0674} & \change{0.1184} & 0.0264 & 0.0620 & 0.3073 & \underline{0.3533} & 0.0288 & 0.0623  \\
GATNE~\cite{cen2019representation} & 0.1137 & 0.1574 & 0.2937 & \underline{0.5046} & 0.0103 & 0.0219 & 0.0153 & 0.0258 & 0.3102 & 0.3426 & \change{0.0055} & \change{0.0124} \\
\midrule
NGCF~\cite{wang2019neural} & \change{0.0659} & \change{0.0917} & \change{0.2538} & \change{0.3960} & \change{0.0090} & \change{0.0670} & \change{0.0063} & \change{0.0098} & \change{0.3038} & \change{0.3347} & \change{0.0061} & \change{0.0136} \\
LightGCN~\cite{he2020lightgcn} & 0.0774 & 0.1093 & 0.2875 & 0.4359 & \change{\underline{0.0757}} & \change{\underline{0.1368}} & 0.0191 & 0.0317 & \change{0.2440} & \change{0.2931} & \underline{0.0362} & \underline{0.0745} \\
MATN~\cite{xia2020multiplex} & \change{0.0920} & \change{0.1417} & \change{0.1595} & \change{0.2999} &\change{0.0193} & \change{0.0612} & \change{0.0244} & \change{0.0446} & \change{0.2098} & \change{0.2394} & \change{0.0062} & \change{0.0135} \\
MB-GMN~\cite{xia2021graph} & \change{0.0906} & \change{0.1763} & \change{0.1179} & \change{0.2456} & \change{0.0413} & \change{0.0944} & \change{0.0430} & \change{0.0724} & \change{0.1583} & \change{0.2120} & \change{0.0047} & \change{0.0116} \\
HybridGNN~\cite{tiankai2022hybridgnn} & 0.1052 & 0.1713 & \underline{0.3020} & 0.4672 & \change{0.0008} & \change{0.0020} & 0.0365 & 0.0612 & 0.1927 & 0.2348 & \change{0.0010} & \change{0.0025} \\
\change{MeLU~\cite{lee2019melu}} & \change{0.1459} & \change{0.2107} & \change{0.2922} & \change{0.3969} & \change{0.0584} & \change{0.1124} & \change{\underline{0.0444}} & \change{0.0666} & \change{0.3082} & \change{0.3249} & \change{0.0244} & \change{0.0536} \\
\midrule
\review{NetWalk~\cite{yu2018netwalk}} & 0.0164 & 0.0690 & 0.0007 & 0.0022 & 0.0003 & 0.0004 & 0.0021 & 0.0030 & 0.0007 & 0.0035 & 0.0002 & 0.0003 \\
DyGNN~\cite{ma2020streaming} & 0.0422 & 0.2547 & 0.0021 & 0.0056 & 0.0032 & 0.0072 & 0.0077 & 0.0214 & 0.0024 & 0.0107 & 0.0003 & 0.0019 \\
EvolveGCN~\cite{pareja2020evolvegcn} & 0.1420 & 0.2271 & \change{0.0071} & \change{0.0118} & 0.0352 & 0.0715 & 0.0414 & \underline{0.0745} & \change{0.0675} & \change{0.1228} & 0.0142 & 0.0287 \\
TGAT~\cite{xu2020inductive} & 0.0758 & 0.1118 & 0.2206 & 0.3304 & 0.0022 & 0.0079 & \change{0.0039} & \change{0.0076} & \change{0.0445} & \change{0.0932} & \change{0.0010} & \change{0.0021} \\
DyHNE~\cite{wang2020dynamic} & \underline{0.2049} & \underline{0.2747} & 0.2354 & 0.4078 &  - & - & 0.0001 &  0.0002 & 0.0046 & 0.0046 & - & - \\
DyHATR~\cite{Xue2020DyHATR} & 0.0565 & 0.1006 & 0.0787 & 0.1652 & \change{0.0012} & \change{0.0030} & \change{0.0032} & \change{0.0063} & \change{0.0107} & \change{0.0189} & \change{0.0009} & \change{0.0027}  \\
\midrule
SUPA & \textbf{0.3312$^{*}$} & \textbf{0.3926$^{*}$} &\textbf{0.3582$^{*}$} & \textbf{0.5107$^{*}$} & \textbf{0.0819$^{*}$} & \textbf{0.1698$^{*}$} & \textbf{0.0471$^{*}$} & \textbf{0.0850$^{*}$} & \textbf{0.3292$^{*}$} & \textbf{0.3575$^{*}$} & \textbf{0.0397$^{*}$} & \textbf{0.0929$^{*}$} \\
\bottomrule
\end{tabular}
\label{tab:exp-link-recall}
\vspace{-10pt}
\end{table*}

\begin{table*}
\scriptsize
\caption{The experimental results of link prediction w.r.t. \textbf{NDCG} and \textbf{MRR}, where ``-'' means that the method fails to give results in a week. The best results are illustrated in bold and the number underlined is the runner-up. Please note that the number with a star ($*$) indicates
\change{the result achieves statistically improvement}
with $p < 0.01$ under t-test compared to other baselines.}
\vspace{-9pt}
\centering
\begin{tabular}{c|cc|cc|cc|cc|cc|cc}
\toprule
& \multicolumn{2}{c|}{UCI} & \multicolumn{2}{c|}{Amazon} & \multicolumn{2}{c|}{Last.fm} & \multicolumn{2}{c|}{MovieLens} & \multicolumn{2}{c|}{Taobao} & \multicolumn{2}{c}{Kuaishou}\\
  & NDCG & MRR & NDCG & MRR & NDCG & MRR & NDCG & MRR & NDCG & MRR & NDCG & MRR \\
\midrule
DeepWalk~\cite{perozzi2014deepwalk} & \change{0.0073} & \change{0.0474} & \change{0.1419} & \change{0.0774} & \change{0.0018} & \change{0.0180} & 0.0210 & 0.0069 & 0.0159 & \underline{0.1764} & 0.0001 & 0.0055 \\
LINE~\cite{tang2015line} & 0.0107 & 0.0249 & \change{0.1665} & \change{0.0661} & \change{0.0005} & \change{0.0112} & \change{0.0302} & \change{0.0071} & \change{0.0194} & \change{0.1758} & \change{0.0001} & \change{0.0031} \\
node2vec~\cite{node2vec-kdd2016} & 0.0155 & 0.0312 & \underline{0.1686} & 0.0619 & \change{0.0013} & \change{0.0181} & 0.0170 & 0.0066 & 0.0142  & 0.1754 & 0.0001 & 0.0073 \\
GATNE~\cite{cen2019representation} & 0.0067 & 0.0347 & 0.1286 & 0.0782 & \underline{0.0032} & 0.0028 & 0.0142 & 0.0052 & \underline{0.0246} & 0.1211 & \change{\underline{0.0008}} & \change{0.0013} \\
\midrule
NGCF~\cite{wang2019neural} & \change{0.0097} & \change{0.0178} & \change{0.0311} & \change{0.0578} & \change{0.0011} & \change{0.0084} & \change{0.0588} & \change{0.0020} & \change{0.0245} & \change{0.1552} & \change{0.0007} & \change{0.0019} \\
LightGCN~\cite{he2020lightgcn} & 0.0116 & 0.0256 & 0.0435 & \underline{0.0818} & \change{0.0004} & \change{\underline{0.0203}} & \underline{0.0778} & 0.0044 & \change{0.0124} & \change{0.1665} & 0.0001 & \underline{0.0093} \\
MATN~\cite{xia2020multiplex} & \change{0.0126} & \change{0.0255} & \change{0.0428} & \change{0.0403} & \change{0.0027} & \change{0.0065} & \change{0.0085} & \change{0.0073} & \change{0.0145} & \change{0.1418} & \change{0.0001} & \change{0.0019} \\
MB-GMN~\cite{xia2021graph} & \change{0.0094} & \change{0.0264} & \change{0.0324} & \change{0.0302} & \change{0.0030} & \change{0.0131} & \change{0.0131} & \change{\underline{0.0257}} & \change{0.0091} & \change{0.1283} & \change{0.0002} & \change{0.0016} \\
HybridGNN~\cite{tiankai2022hybridgnn} & 0.0078 & 0.0314 &  0.0680 & 0.0692 & \change{0.0014} & \change{0.0003} & 0.0234 & 0.0220 & 0.0226 & 0.0806 & \change{0.0003} & \change{0.0004} \\
\change{MeLU~\cite{lee2019melu}} & \change{0.0178} & \change{0.0382} & \change{0.1619} & \change{0.0753} & \change{0.0003} & \change{0.0157} & \change{0.0157} & \change{0.0124} & \change{0.0158} & \change{0.1645} & \change{0.0001} & \change{0.0064} \\
\midrule
\review{NetWalk~\cite{yu2018netwalk}} & 0.0027 & 0.0091 & 0.0018 & 0.0006 & 0.0004 & 0.0001 & 0.0145 & 0.0007 & 0.0001 & 0.0008 & 0.0001 & 0.0001 \\
DyGNN~\cite{ma2020streaming} & 0.0054 & 0.0342 & 0.0001 & 0.0009 & 0.0015 & 0.0011 & 0.0194 & 0.0025 & 0.0004 & 0.0024 & 0.0005 & 0.0003 \\
EvolveGCN~\cite{pareja2020evolvegcn} & \underline{0.0201} & 0.0342 & \change{0.0091} & \change{0.0034} & 0.0017 & 0.0064 & 0.0082 & 0.0102 & \change{0.0022} & \change{0.0197} & 0.0002 & 0.0027 \\
TGAT~\cite{xu2020inductive} & 0.0036 & 0.0255 & 0.0779 & 0.0514 & 0.0025 & 0.0012 & \change{0.0108} & \change{0.0018} & \change{0.0025} & \change{0.0122} & \change{\underline{0.0008}} & \change{0.0002} \\
DyHNE~\cite{wang2020dynamic} & 0.0089 & \underline{0.0568} & 0.0429 & 0.0566 & - & -  & 0.0001 & 0.0003 & 0.0002 & 0.0027 & - & - \\
DyHATR~\cite{Xue2020DyHATR} & 0.0030 & 0.0124 & 0.0065 & 0.0182 & \change{0.0014} & \change{0.0006} & \change{0.0061} & \change{0.0010} & \change{0.0017} & \change{0.0029} & \change{0.0006} & \change{0.0004}  \\
\midrule
SUPA & \textbf{0.0215$^{*}$} & \textbf{0.1174$^{*}$} & \textbf{0.1750$^{*}$} & \textbf{0.0849$^{*}$} & \textbf{0.0045$^{*}$} & \textbf{0.0234$^{*}$} & \textbf{0.0918$^{*}$} & \textbf{0.0273$^{*}$} & \textbf{0.0277$^{*}$} & \textbf{0.1906$^{*}$} & \textbf{0.0010$^{*}$} & \textbf{0.0122$^{*}$} \\
\bottomrule
\end{tabular}
\label{tab:exp-link-mrr}
\vspace{-10pt}
\end{table*}

We use six datasets to analyze the performance of SUPA, namely UCI, Amazon, Last.fm, MovieLens, Taobao, and Kuaishou.
\change{The statistics and the selected multiplex schemas of these datasets are summarized in Table~\ref{tab:datasets} and Table~\ref{tab:metapaths}, respectively.
The detailed information of these datasets is listed as follows.}
\begin{itemize}
    \item \textbf{UCI}~\cite{konect:2016:opsahl-ucsocial} contains the message communications between the students of the University of California, Irvine in an online community. It is a streaming homogeneous network
    and we use the data provided by~\cite{ma2020streaming}. \change{Note that $\mathcal{O} = $\{\textbf{U}ser\}, $\mathcal{R} =$ \{\textbf{C}ommunicate\}.}
    \item \textbf{Amazon}~\cite{he2016ups} is a static multiplex heterogeneous network that includes product metadata and links between products. We use the data provided by~\cite{cen2019representation} which contains the product metadata of \textit{Electronic} category.
    \change{Note that $\mathcal{O} =$ \{\textbf{P}roduct\}, $\mathcal{R} =$ \{\textbf{L}ink\}.}
    \item \textbf{Last.fm}\footnote{https://www.last.fm/} contains $<$user, timestamp, artist, song$>$ tuples collected from Last.fm API, which represents the whole listening habits (till May, 5th 2009) of nearly 1,000 users. We use the $<$user, artist$>$ pairs with the first interactive timestamp to generate a heterogeneous network.
    \change{Note that $\mathcal{O} =$ \{\textbf{U}ser, \textbf{A}rtist\}, $\mathcal{R} =$ \{\textbf{L}isten to\}.}
    \item \textbf{MovieLens}~\cite{10.1145/2827872} describes 5-star rating and free-text tagging activity between January 09, 1995 and March 31, 2015 from MovieLens, a movie recommendation service.
    \change{Note that $\mathcal{O} =$ \{\textbf{U}ser, \textbf{M}ovie\}, $\mathcal{R}$ = \{\textbf{R}ate, \textbf{T}ag\}.}
    \item \textbf{Taobao}~\cite{zhu2018learning} is offered by Alibaba with user behaviors collected from Taobao\footnote{https://www.taobao.com/}.
    There are 1,000 users with all the corresponding interactive items in this dataset.
    \change{Note that $\mathcal{O} =$ \{\textbf{U}ser, \textbf{I}tem\}, $\mathcal{R} =$ \{\textbf{P}age view, \textbf{B}uy, \textbf{C}art, \textbf{F}avorite\}.}
    \item \textbf{Kuaishou}\footnote{https://www.kuaishou.com/} is collected from the Kuaishou online video-watching platform. This dataset includes the interactions of 6,840 users and 131,972 videos in a week of time. 
    \change{Note that $\mathcal{O} =$ \{\textbf{U}ser, \textbf{V}ideo, \textbf{A}uthor\}, $\mathcal{R} =$ \{\textbf{W}atch, \textbf{L}ike, \textbf{F}orward, \textbf{C}omment, \textbf{U}pload\}.}
\end{itemize}

\enlargethispage{2em}
\todo{Note that Amazon is a static graph while UCI and Last.fm are not multiplex. We use them to evaluate the generalization ability of SUPA, i.e., to evaluate whether SUPA can handle static networks and homogeneous networks well.}
\subsection{Baseline Methods}
To demonstrate the effectiveness and efficiency of SUPA, we choose \change{sixteen} state-of-the-art baseline methods, categorized into three groups.
Methods for static network embedding include DeepWalk~\cite{perozzi2014deepwalk}, LINE~\cite{tang2015line}, node2vec~\cite{node2vec-kdd2016} and GATNE~\cite{cen2019representation}.
Methods for recommendation include NGCF~\cite{wang2019neural}, LightGCN~\cite{he2020lightgcn}, MATN~\cite{xia2020multiplex}, MB-GMN~\cite{xia2021graph}, HybridGNN~\cite{tiankai2022hybridgnn} and \change{MeLU~\cite{lee2019melu}}.
Methods for dynamic network embedding include \todo{NetWalk~\cite{yu2018netwalk}}, DyGNN~\cite{ma2020streaming}, EvolveGCN~\cite{pareja2020evolvegcn}, TGAT~\cite{xu2020inductive}, DyHNE~\cite{wang2020dynamic} and DyHATR~\cite{Xue2020DyHATR}.
The details of the baseline methods are listed as follows.
\subsubsection{\todo{Methods for static network embedding}}
\begin{itemize}
    \item \textbf{DeepWalk}~\cite{perozzi2014deepwalk} is an embedding method for static homogeneous networks. It exploits the random walk strategy and the skip-gram model to learn node vector representations, which ignores the heterogeneous and temporal information during training.
    \item \textbf{LINE}~\cite{tang2015line} is also a static homogeneous network embedding method.
    It learns node representations by modelling the first- and second-order proximity between node pairs. LINE is trained similar to DeepWalk.
    \item \textbf{node2vec}~\cite{node2vec-kdd2016} adds two parameters to control the random walk process based on DeepWalk. It is also designed for static homogeneous networks.
    \item \textbf{GATNE}~\cite{cen2019representation} is a state-of-the-art GNN model for static multiplex heterogeneous networks. It learns different edge embeddings for nodes. Notice that GATNE is not able to model temporal information.
\end{itemize}
\subsubsection{\todo{Methods for recommendation}}
\begin{itemize}
    \item \textbf{NGCF}~\cite{wang2019neural} is a message passing architecture for information aggregation over the user-item interaction graph, to exploit high-order connection relationships.
    \item \textbf{LightGCN}~\cite{he2020lightgcn} is a light-weight graph convolution network, which is easy to train and has good generalization ability.
    \enlargethispage{0.5em}
    \item \textbf{MATN}~\cite{xia2020multiplex} differentiates the relations between users and items with the integration of the attention network and memory units.
    \item \textbf{MB-GMN}~\cite{xia2021graph} is a multi-behavior enhanced recommendation framework with graph meta network for learning interaction heterogeneity and diversity.
    \item \textbf{HybridGNN}~\cite{tiankai2022hybridgnn} uses hybrid aggregation flows and hierarchical attentions to fully utilize the heterogeneity in the multiplex scenarios to learn representations for recommendation.
    \item \change{\textbf{MeLU}~\cite{lee2019melu} is a meta-learning based recommender system which can estimate a new user’s preferences based on only a few user-item interactions.}
\end{itemize}
\subsubsection{\todo{Methods for dynamic network embedding}}
\begin{itemize}
    \item \todo{\textbf{NetWalk~\cite{yu2018netwalk}} adopts clique embedding and a deep autoencoder to encode the nodes into vectors, which can be updated dynamically as the network evolves.}
    \item \textbf{DyGNN}~\cite{ma2020streaming} is a dynamic GNN model for homogeneous network. It is designed for steaming networks and is able to capture the information hidden in time intervals.
    \enlargethispage{0.5em}
    \item \textbf{EvolveGCN}~\cite{pareja2020evolvegcn} is a convolutional network for Dynamic Graphs. It captures the dynamics of the graph sequence by using an RNN to evolve the GCN parameters. 
    \item \textbf{TGAT}~\cite{xu2020inductive} utilizes a temporal graph attention layer to aggregate temporal-topological neighborhood features and learn the time-feature interactions.
    \item \textbf{DyHNE}~\cite{wang2020dynamic} is a dynamic heterogeneous embedding model, which introduces the metapath-based first- and second-order proximity to preserve structural and semantic information. 
    \item \textbf{DyHATR}~\cite{Xue2020DyHATR} uses hierarchical attention to learn heterogeneous information and incorporates RNN with temporal attention to capture evolutionary patterns.
\end{itemize}

\begin{figure*}
\centering
\begin{minipage}{0.21\linewidth}
\centering
\includegraphics[width=1\linewidth]{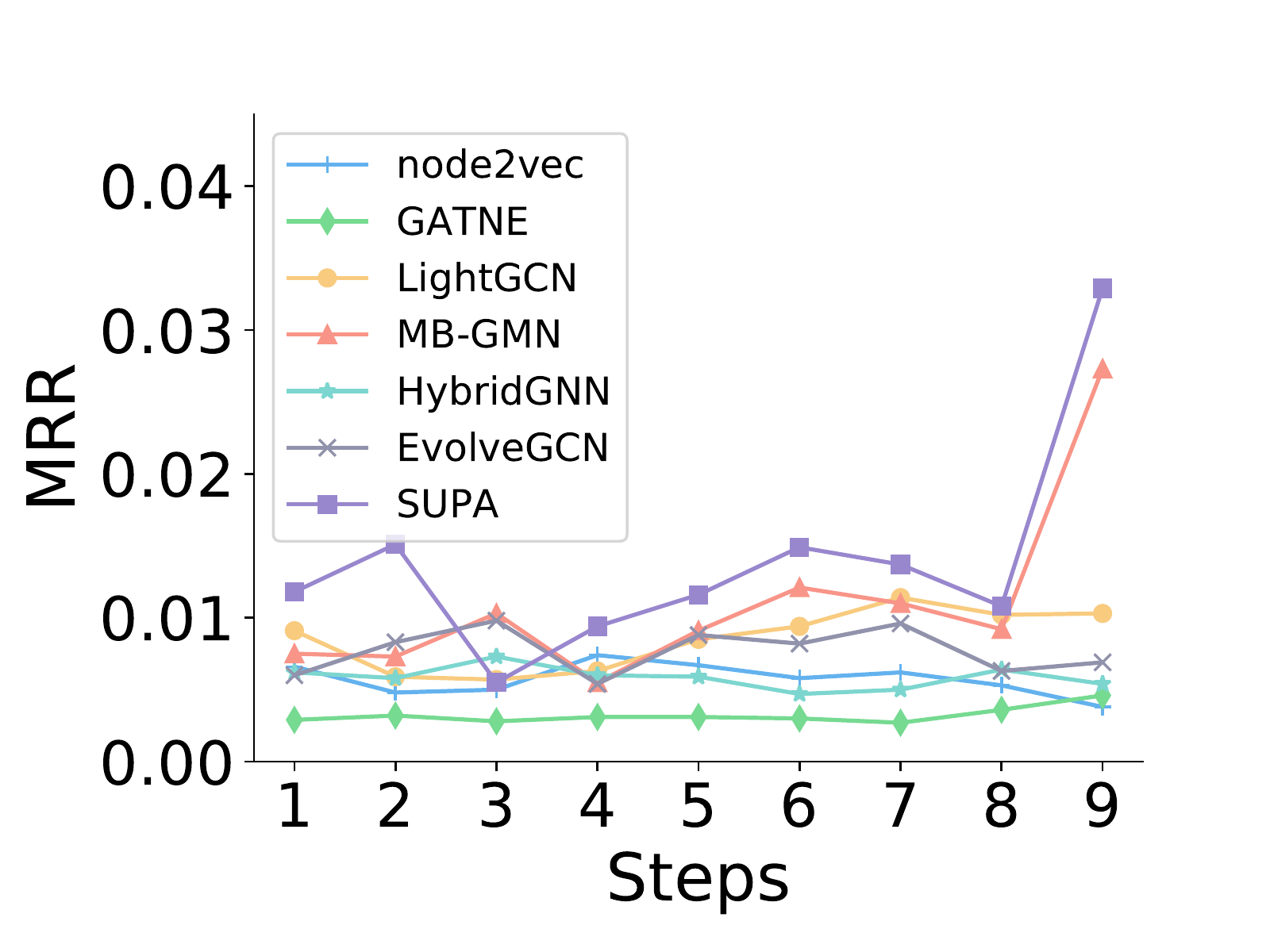}
\vspace{-20pt}
\caption{The experimental results of dynamic link prediction on MovieLens.}
  \label{fig:dynamic_lp}
\end{minipage}
\quad
\begin{minipage}{0.21\linewidth}
\centering
\includegraphics[width=1\linewidth]{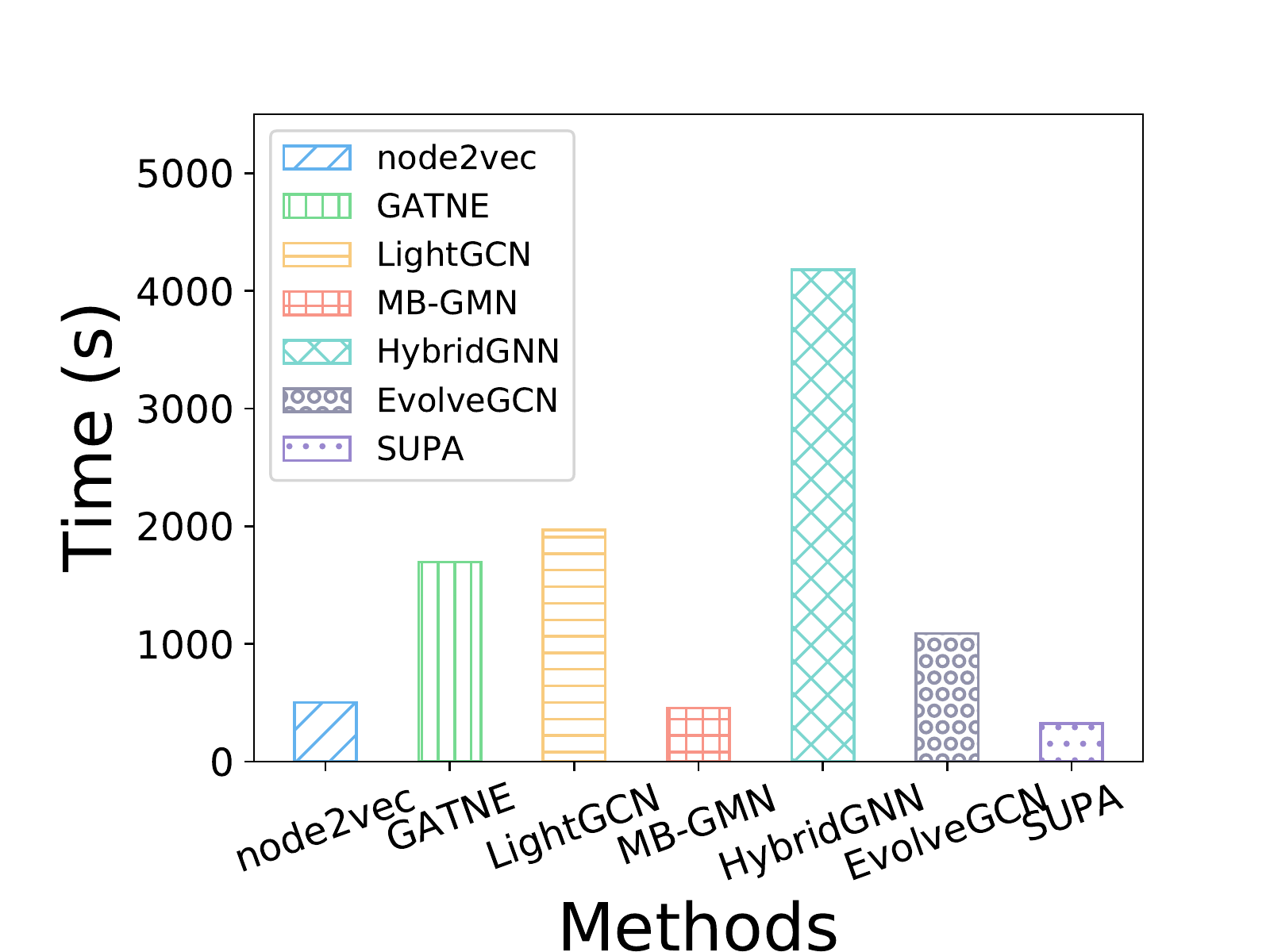}
\vspace{-20pt}
\caption{\todo{The running time of dynamic link prediction on MovieLens.}}
\label{fig:dynamic_lp_time}
\end{minipage}%
\quad
\begin{minipage}{0.21\linewidth}
\centering
\includegraphics[width=1\linewidth]{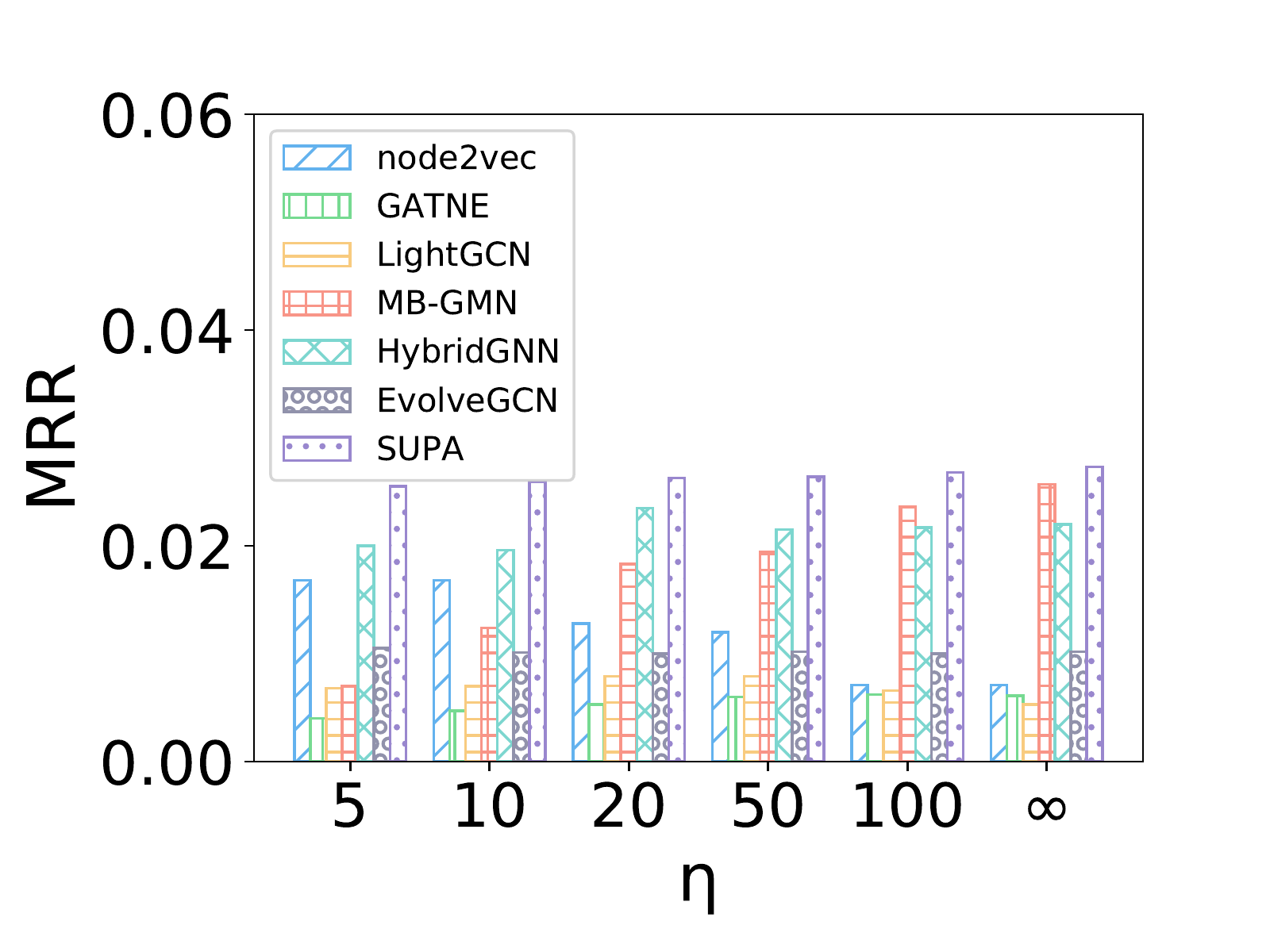}
\vspace{-20pt}
\caption{The experimental results of robustness to neighborhood disturbance on MovieLens.}
\label{fig:robustness}
\end{minipage}
\quad
\begin{minipage}{0.21\linewidth}
\centering
\includegraphics[width=1\linewidth]{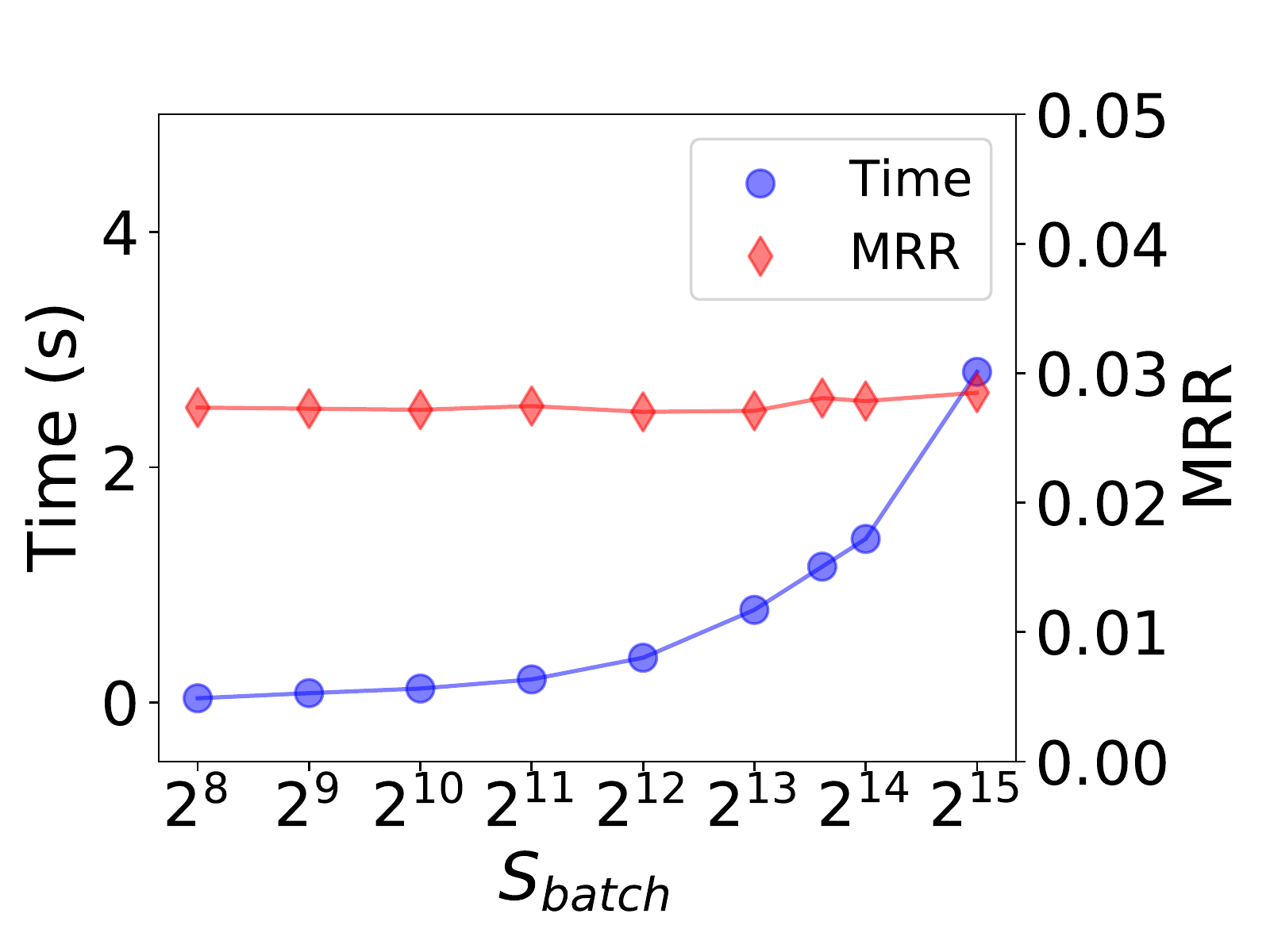}
\vspace{-20pt}
\caption{\change{The experimental results of Scalability.}}
\label{fig:scalability}
\end{minipage}
\end{figure*}

\subsection{Experimental Settings}
For the hyper-parameters in SUPA, we use grid search to obtain the optimal hyper-parameters. Specifically, we set the number of sampling paths $n$ from $1$ to $20$, the path length $l$ from $1$ to $10$ and the negative node numbers as $[1,3,5,7]$. We calculate $\tau$ from $g(\tau) = 1 / \log(e + \tau) = 0.3$.
For the hyper-parameters in the training workflow,
we set $S_{batch}$ as $1024$, $I_{valid}$ to $8$, $S_{valid}$ to $150$ and $\mu$ to $3$. The $N_{iter}$ is set to $100$ for the UCI and Taobao datasets, and to $30$ for the others.
We use the Adam optimizer to train our model with an initial learning rate $3e^{-3}$ and weight decay $1e^{-4}$.
For all the datasets, we take $80\%$ of the edges as training set, $1\%$ as validation set and the rest $19\%$ as test set.
\todo{To evaluate on the static dataset (i.e., Amazon), we train methods for dynamic networks by assuming that all the edges share a same timestamp.}
For all the baseline methods, we tune the parameters on the validation set. The node embedding size of all the model is set to $128$, and all the experiments are conducted on a single GTX 1080Ti GPU.

\enlargethispage{2em}
\todo{
To evaluate the performance of each model, we choose three widely used metrics, namely \textbf{H(it rate)@k, normalized discounted cumulative gain (\textbf{NDCG@K})~\cite{xia2021graph} and mean reciprocal rank (\textbf{MRR})~\cite{voorhees1999trec, ma2020streaming}}.
\begin{itemize}
    \item \textbf{H@K} calculates the rate that the ground truth nodes are ranked in top K out of all the nodes, and higher \textbf{H@K} indicates better performance. We report \textbf{H@20} and \textbf{H@50} in this paper.
    \item \textbf{NDCG} assigns higher scores to hits at a higher position in the top-K ranking list, which emphasizes that test items should be ranked as higher as possible. We report $K = 10$ in this paper due to page limit. Notice that similar results are observed when $K=20$ and $K=50$.
    \item \textbf{MRR} calculates the mean of the reciprocal ranking of the ground truth nodes in the testing set. The higher \text{MRR} is, the more ground truth nodes are ranked top out of all the nodes.
\end{itemize}
}


\begin{table*}
\scriptsize
\vspace{-10pt}
\caption{The experimental results of ablation study w.r.t. the contribution of different kinds of losses \change{and the effectiveness of the InsLearn training flow}.}
\vspace{-5pt}
\label{tab:losses}
\centering
\begin{tabular}{c|cc|cc|cc|cc|cc|cc}
\toprule
& \multicolumn{2}{c|}{UCI} & \multicolumn{2}{c|}{Amazon} & \multicolumn{2}{c|}{Last.fm} & \multicolumn{2}{c|}{MovieLens} & \multicolumn{2}{c|}{Taobao} & \multicolumn{2}{c}{Kuaishou}\\
  & H@50 & MRR & H@50 & MRR & H@50 & MRR & H@50 & MRR & H@50 & MRR & H@50 & MRR \\
\midrule
SUPA$_{\mathcal{L}_{inter}}$ & 0.2680 & 0.0209 & 0.5094 & 0.0844 & 0.1478 & 0.0150 & \textbf{0.1019} & \textbf{0.0305} & 0.1977 & 0.1134 & 0.0228 & 0.0036 \\
SUPA$_{\mathcal{L}_{prop}}$ & 0.2249 & 0.0341 & 0.2839 & 0.0514 & 0.1315 & 0.0156 & 0.0600 & 0.0084 & 0.3192 & 0.1635 & 0.0308 & 0.0041 \\
SUPA$_{\mathcal{L}_{neg}}$ & 0.1907 & 0.0188 & 0.0504 & 0.0076 & 0.1144 & 0.0113 & 0.0753 & 0.0259 & 0.0226 & 0.0032 & 0.0201 & 0.0028 \\
\midrule
SUPA$_{w/o \mathcal{L}_{inter}}$ & 0.3603 & 0.1130 & 0.5040 & 0.0846 & 0.1627 & 0.0220 & 0.0778 & 0.0250 & 0.3295 & 0.1805 & 0.0906 & 0.0120 \\
SUPA$_{w/o \mathcal{L}_{prop}}$ & 0.2528 & 0.0239 & 0.0683 & 0.0104 & 0.1204 & 0.0115 & 0.0791 & 0.0249 & 0.0211 & 0.0914 & 0.0196 & 0.0028 \\
SUPA$_{w/o \mathcal{L}_{neg}}$ & 0.2237 & 0.0325 & 0.2813 & 0.0514 & 0.1329 & 0.0155 & 0.0976 & 0.0294 & 0.3029 & 0.1616 & 0.0303 & 0.0040 \\
\midrule
\change{SUPA$_{w/o Ins}$} & \change{0.3621} & \change{0.0999} & \change{\textbf{0.5257}} & \change{0.0786} & \change{0.1395} & \change{0.0206} & \change{0.0742} & \change{0.0236} & \change{0.3570} & \change{0.1787} & \change{0.0778} & \change{0.0101} \\
\midrule
SUPA & \textbf{0.3926} & \textbf{0.1174} & 0.5107 & \textbf{0.0849} & \textbf{0.1698} & \textbf{0.0234} & 0.0850 & 0.0273 & \textbf{0.3575} & \textbf{0.1906} & \textbf{0.0929} & \textbf{0.0122} \\
\bottomrule
\end{tabular}
\vspace{-15pt}
\end{table*}

\subsection{Link Prediction}
\label{sec:link_pred}

We evaluate the performance of SUPA on the link prediction task  \todo{(i.e., making recommendation)}. Based on the results reported in Table~\ref{tab:exp-link-recall} and Table~\ref{tab:exp-link-mrr}, we have the following observations:
1) SUPA consistently performs the best on all datasets compared to other methods. We attribute these results to the fact that SUPA provides components to capture the multiplex heterogeneity and streaming dynamics.
2) Although SUPA is designed for multiplex heterogeneous networks, it outperforms all the baselines in the homogeneous network (i.e., UCI), the static networks (i.e., Amazon) and the non-multiplex heterogeneous network (i.e., Last.fm), which illustrates the robustness and generalization ability of SUPA.
3) DeepWalk and node2vec are not neighbor-aggregation based methods and are free of neighborhood disturbance, which makes them have better generalization ability than other baseline methods.
4) Methods for recommendation give comparable results with SUPA on some datasets, but they are not able to benefit from temporal information.
5) Dynamic network embedding methods have poor performances at most cases because they cannot capture the complex interactive patterns between users and items and then may be not suitable for real challenging scenarios, such as recommender systems for massive short videos.
Besides, DyHNE requires time-consuming matrix computation and cannot produce results in a week on Last.fm and Kuaishou.
Note that node2vec, GATNE, LightGCN, MB-GMN, \todo{HybridGNN} and Evolve-GCN have better performances than other baseline methods, and thus we select them as baseline methods in Section~\ref{sec:exp_dlp} and Section~\ref{sec:exp_nd}. Besides, we observe that all the metrics share similar tendency, and thus we only report H@50 and MRR in the following experiments.

\enlargethispage{2em}
\subsection{Dynamic Link Prediction}
\label{sec:exp_dlp}
In this subsection, we evaluate the practicability of different methods for real-world recommender systems, where recommendation models are required to repeatedly learn from new data and preform dynamic link prediction. Specifically, we sort the edge set  $\mathcal{E}$ according to the time information and split $\mathcal{E}$ into 10 parts with equal size, denoted as $\{\mathcal{E}_1,\mathcal{E}_2,\cdots, \mathcal{E}_{10} \}$. Then, each method is required to retrain (i.e., for methods dealing with static networks) or incrementally train (i.e., for methods dealing with dynamic networks) on $\mathcal{E}_i$ and evaluate on $\mathcal{E}_{i+1}$ ($i = 1, 2, \cdots, 9$). We conduct the experiment on the MovieLens dataset and evaluate SUPA with node2vec, GATNE, LightGCN, MB-GMN, \todo{HybridGNN} and EvolveGCN.

As shown in Figure~\ref{fig:dynamic_lp}, SUPA has the best prediction results in most cases due to the carefully design to handle time information and the incrementally learning ability thanks to InsLearn.
MB-GMN is the best baseline method since it is designed for recommendation on multiplex heterogeneous networks. However, MB-GMN fails to benefit from the temporal information.
In contrast, EvolveGCN can utilize the time information and obtains good performance in the first five steps. Nevertheless, EvolveGCN has no components to deal with out-date information and suffers from noises from historical data in the last four steps.
Note that there is a long time interval
between Step 2 and Step 3. The user preferences may change a lot during this period. As a result, SUPA has a slight performance drawback since SUPA remembers some user preferences using the long-term memories. In addition, at Step 9, methods designed to handle multiplex heterogeneous networks (i.e., SUPA and MB-GMN) obtain excellent performances. The reason is that in $\mathcal{E}_{10}$, there are many interactions happened between nodes that have established edges with a different relation in $\mathcal{E}_9$, which can be captured by SUPA and MB-GMN.
\review{HybridGNN fails to benefit from multiplex heterogeneity because the graph w.r.t. each $\mathcal{E}_i$ is too sparse for HybridGNN to form high-quality aggregation flows.}

We also report the sum of the running time of different methods in Figure~\ref{fig:dynamic_lp_time}. It can be observed that SUPA is more efficient than other methods, indicating the model architecture of SUPA and the InsLearn training workflow can not only improve the effectiveness of recommendation, but also shorten the inference time.

\subsection{Robustness to Neighborhood Disturbance}
\label{sec:exp_nd}
In this subsection, we run the link prediction task in a real-world situation where only the most recent subgraph is available due to resource constraints to validate the robustness of SUPA to neighborhood disturbance.
Specifically, we denote the maximum number of neighbors preserved for each node in the graph as $\eta$.
Then, different methods are trained on the subgraph where each node only has the latest $\eta$ neighbors at each time step.
Note that methods for static graphs can only see the subgraph at the final time step.
We choose node2vec, GATNE, LightGCN, MB-GMN, \todo{HybridGNN}, EvolveGCN as the baseline methods, and vary $\eta$ in $\{5, 10, 20, 50, 100, \infty\}$.

\enlargethispage{2em}
The link prediction results on the MovieLens dataset are shown in Figure~\ref{fig:robustness}.
SUPA obtains the best performances and is insensitive to $\eta$ since the propagation-based neural network architecture of SUPA can effectively deal with neighborhood disturbance.
EvolveGCN is also insensitive to $\eta$ since it is a dynamic network embedding method and is aware of new neighbors of nodes.
Other baseline methods have different behaviors as $\eta$ varies.
This is because subgraphs with smaller $\eta$ highlight the short-term interest of users while subgraphs with larger $\eta$ maintain users' long-term preferences, and thus a balance is required for these baselines.

\begin{table}
\scriptsize
\caption{The experimental results of ablation study w.r.t. the benefits of modeling multiplex heterogeneity and streaming dynamics.}
\vspace{-5pt}
\label{tab:benefit}
\centering
\begin{tabular}{c|cc|cc}
\toprule
& \multicolumn{2}{c|}{Taobao} & \multicolumn{2}{c}{Kuaishou} \\
  & H@50 & MRR & H@50 & MRR  \\
\midrule
SUPA$_{sn}$ & 0.3417 & 0.1801 & 0.0743 & 0.0099 \\
SUPA$_{se}$ & 0.3468 & 0.1844 & 0.0745 & 0.0100 \\
SUPA$_{s}$ & 0.3209 & 0.1772 & 0.0635 & 0.0087 \\
\midrule
SUPA$_{nf}$ & 0.3472  & 0.1856 & 0.0744 & 0.0100 \\
SUPA$_{nd}$ & 0.3474 & 0.1854 & 0.0742 & 0.0100 \\
SUPA$_{nt}$ & 0.3472  & 0.1827 & 0.0667 & 0.0091 \\
\midrule
SUPA & \textbf{0.3575}  & \textbf{0.1906} & \textbf{0.0929} & \textbf{0.0122} \\
\bottomrule
\end{tabular}
\end{table}

\begin{figure*}[ht]
  \centering
  \subfigure[Impact of $d$]{
    \begin{minipage}[]{0.17\linewidth}
    \includegraphics[width=1.0\linewidth]{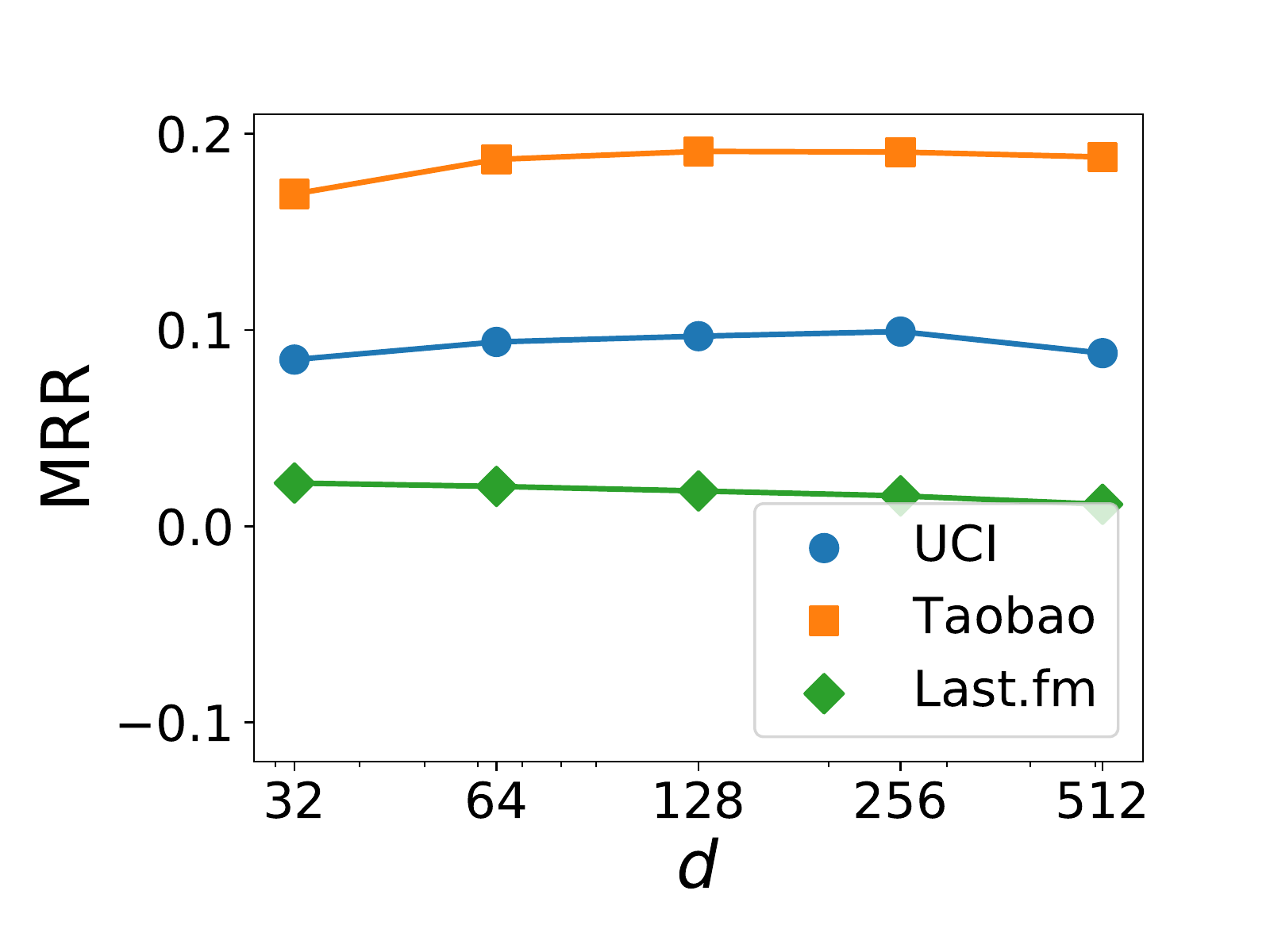}
    \end{minipage}
  }
  \subfigure[Impact of $k$]{
    \begin{minipage}[]{0.17\linewidth}
  \includegraphics[width=1.0\linewidth]{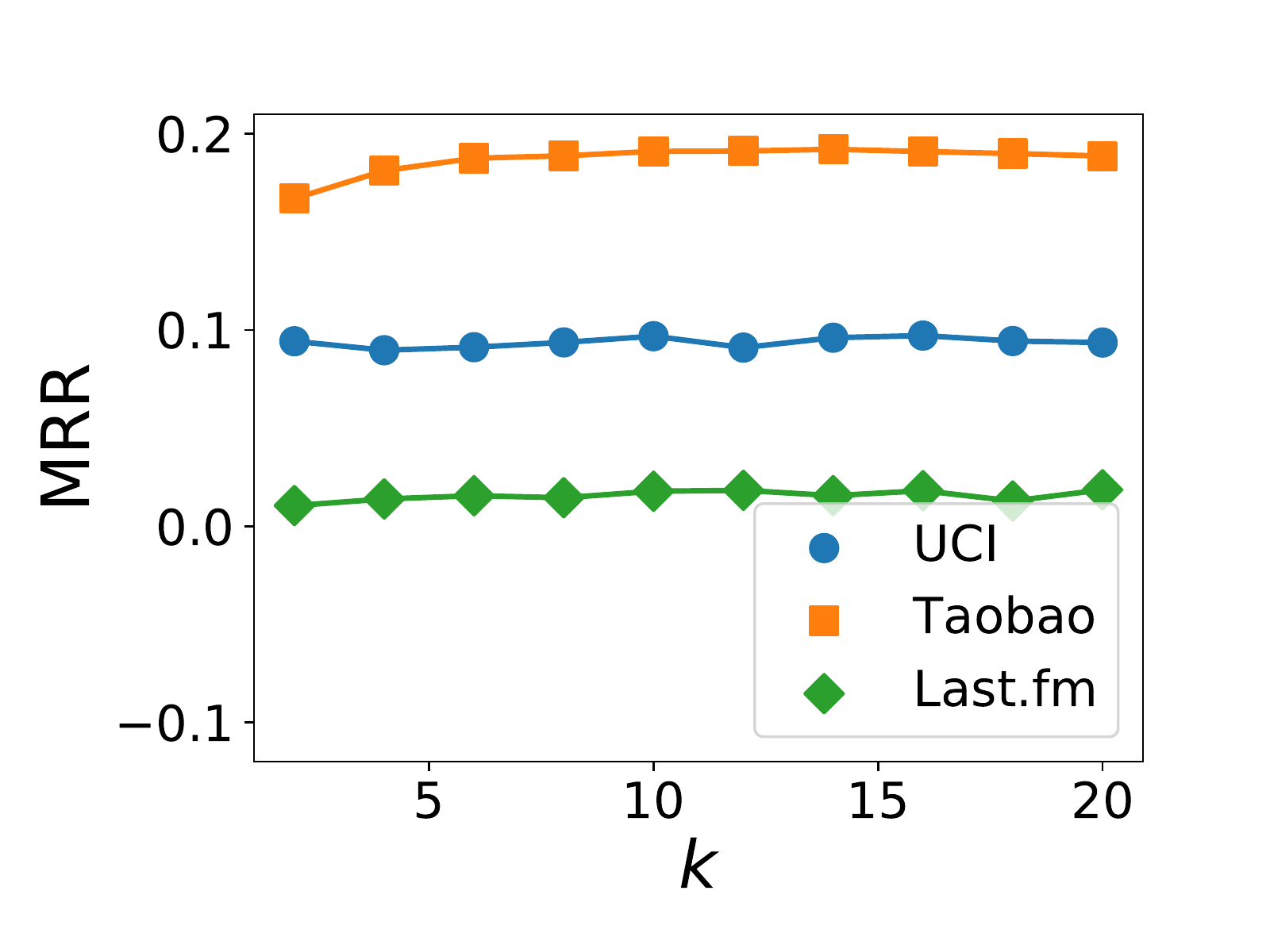}
    \end{minipage}
  }
  \subfigure[Impact of $l$]{
    \begin{minipage}[]{0.17\linewidth}
  \includegraphics[width=1.0\linewidth]{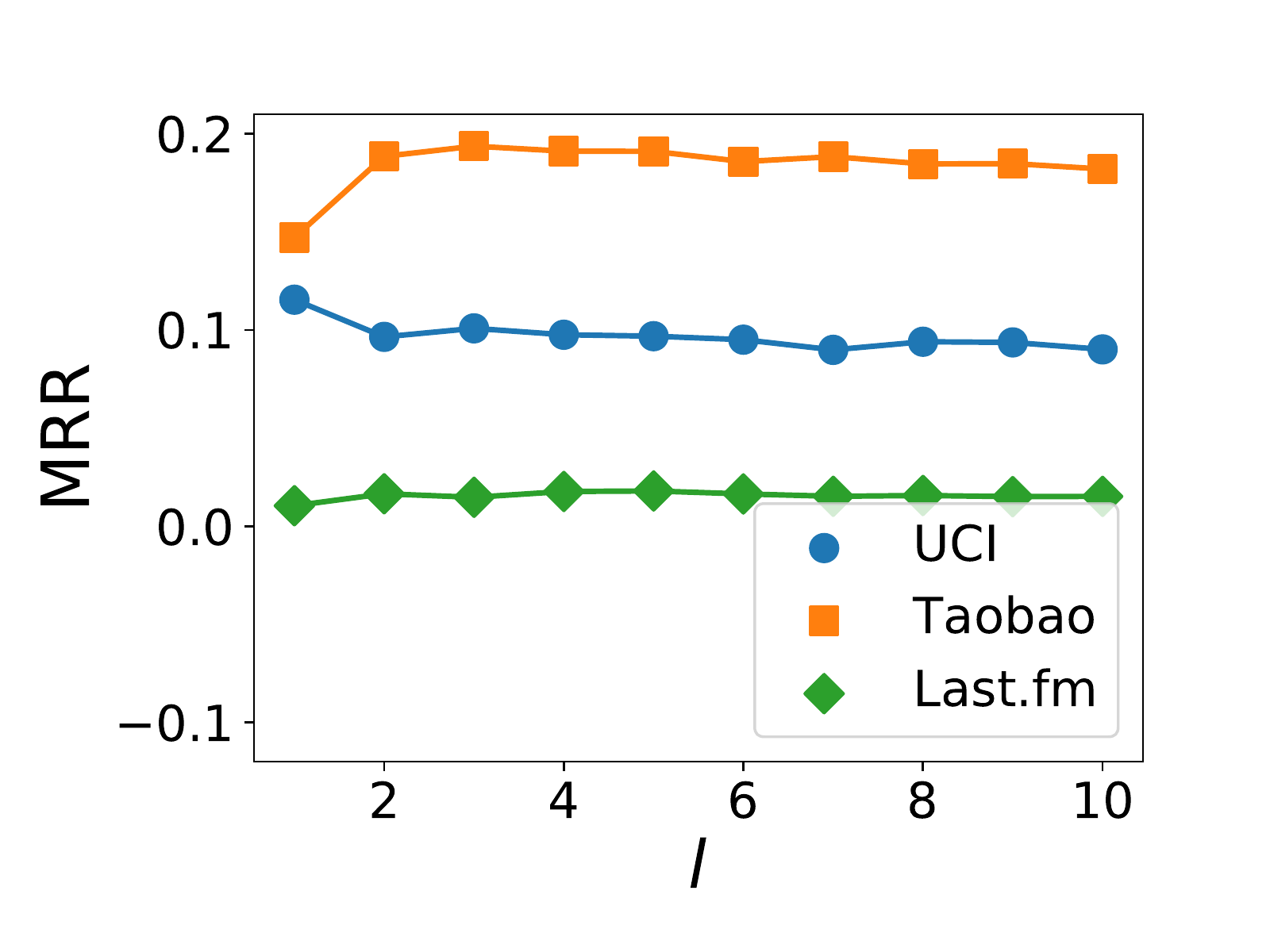}
    \end{minipage}
  }
  \subfigure[Impact of $N_{neg}$]{
    \begin{minipage}[]{0.17\linewidth}
  \includegraphics[width=1.0\linewidth]{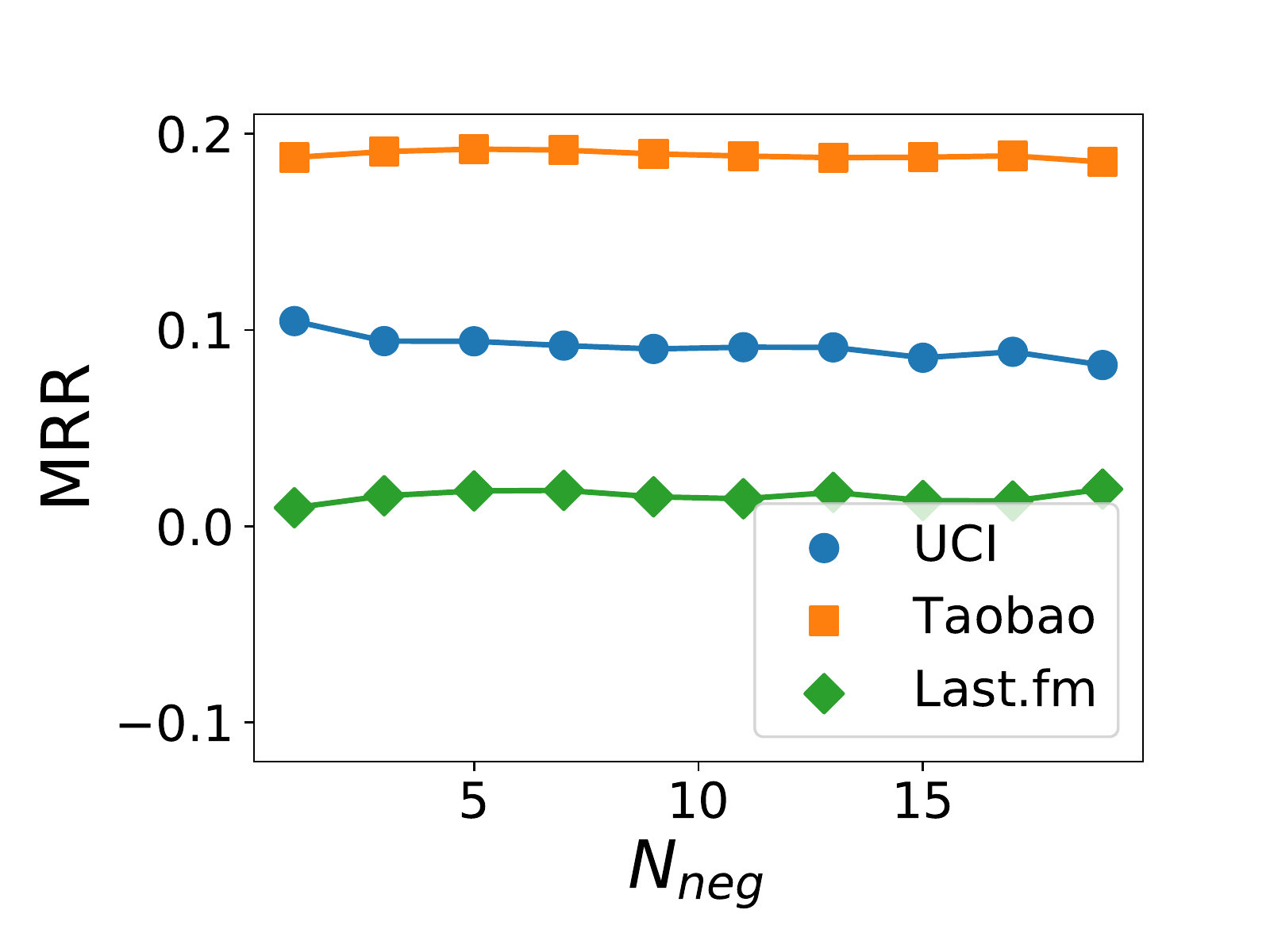}
    \end{minipage}
  }
  \subfigure[Impact of $\tau$]{
    \begin{minipage}[]{0.17\linewidth}
  \includegraphics[width=1.0\linewidth]{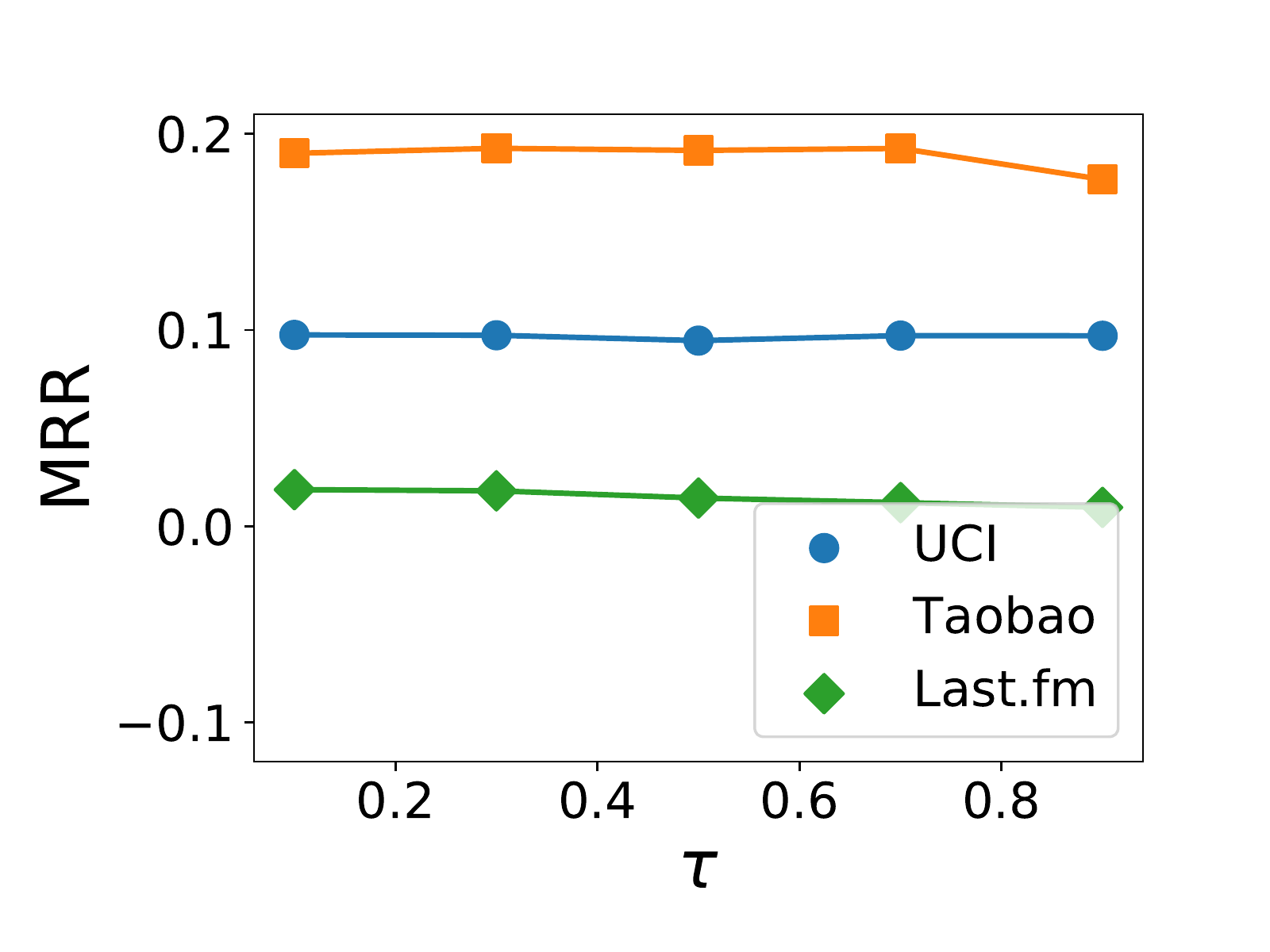}
    \end{minipage}
  }
  \subfigure[Impact of $N_{iter}$]{
    \begin{minipage}[]{0.17\linewidth}
    \includegraphics[width=1.0\linewidth]{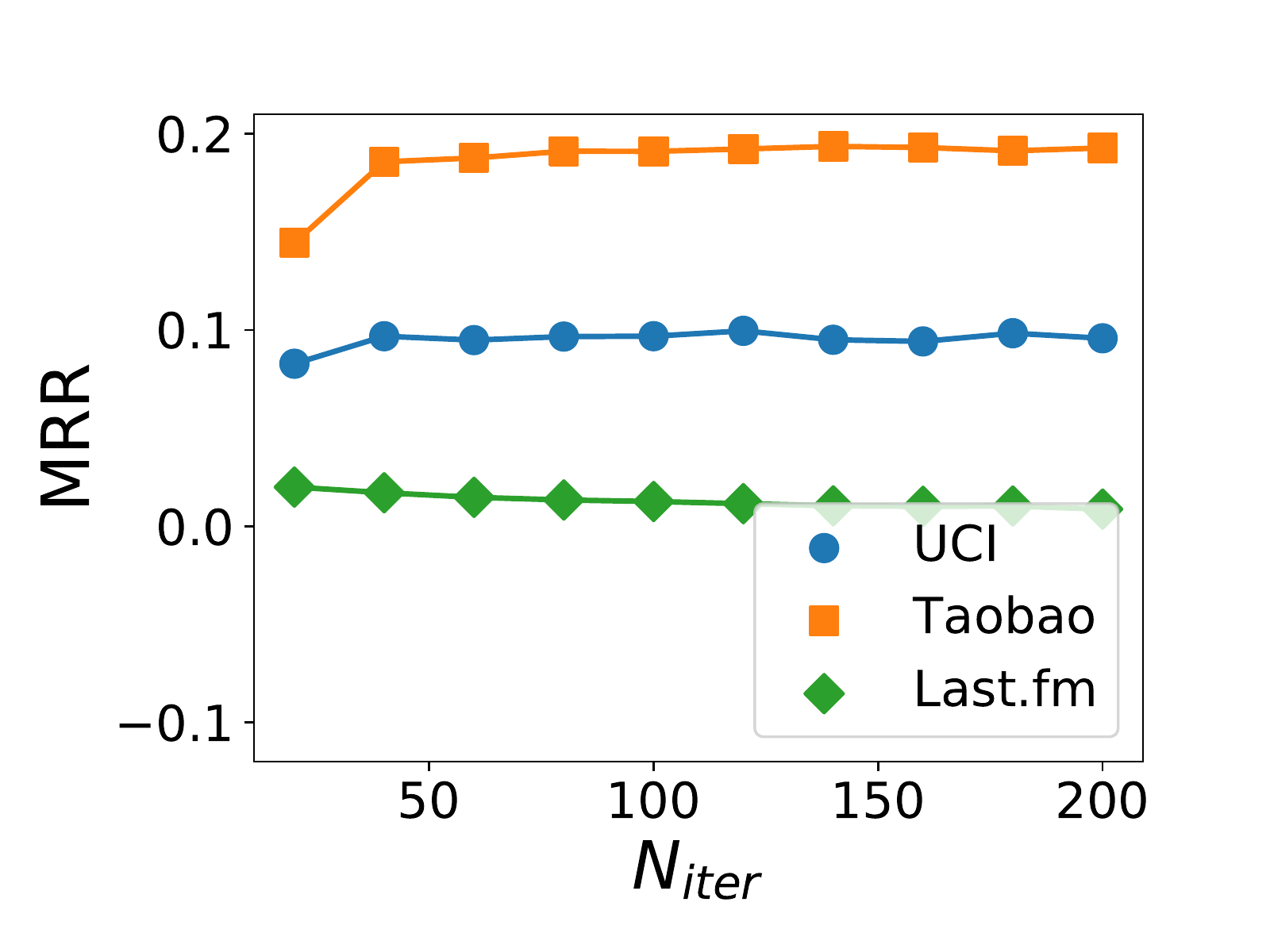}
    \end{minipage}
  }
  \subfigure[Impact of $I_{valid}$]{
    \begin{minipage}[]{0.17\linewidth}
  \includegraphics[width=1.0\linewidth]{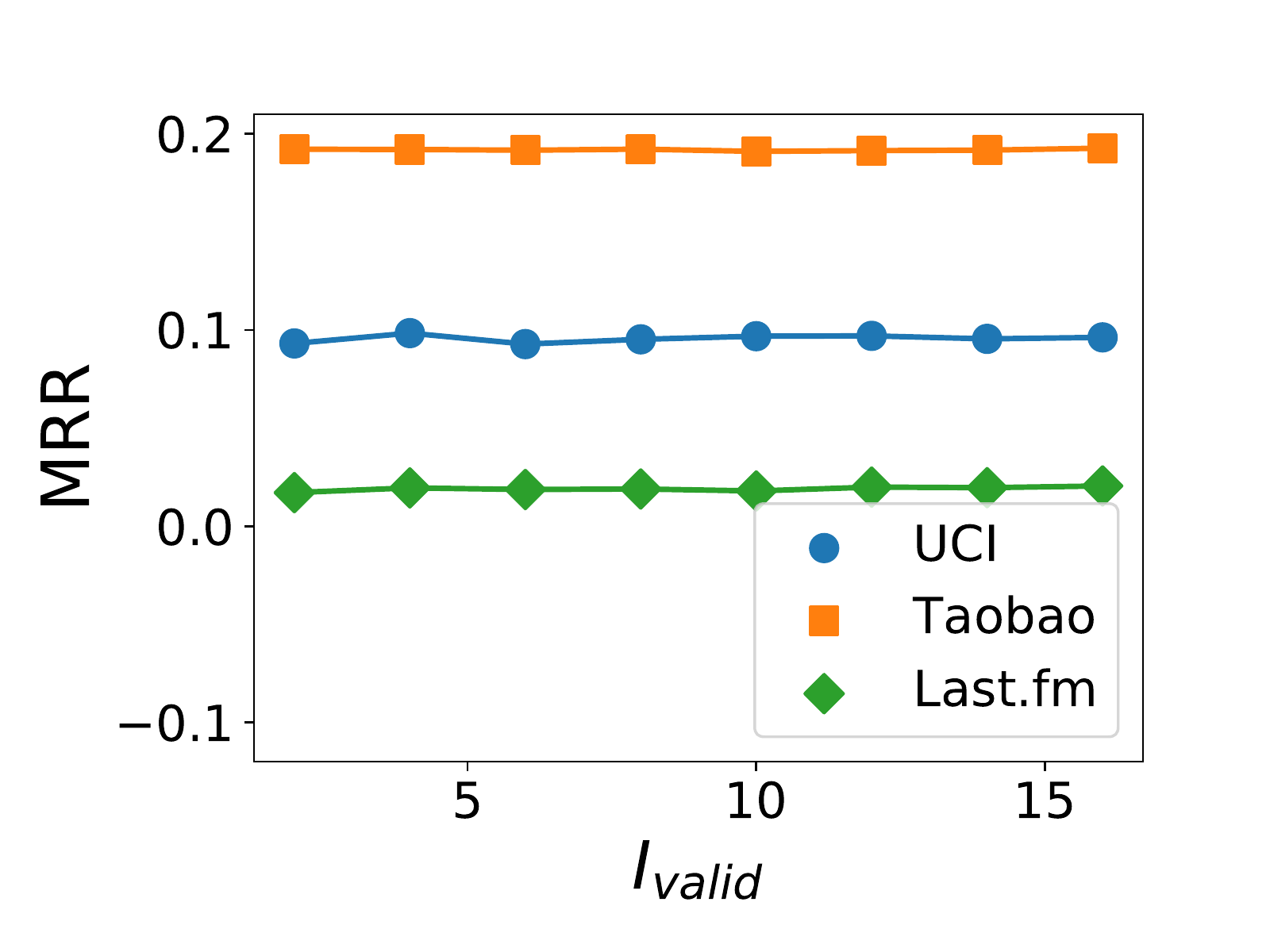}
    \end{minipage}
  }
  \subfigure[Impact of $S_{valid}$]{
    \begin{minipage}[]{0.17\linewidth}
  \includegraphics[width=1.0\linewidth]{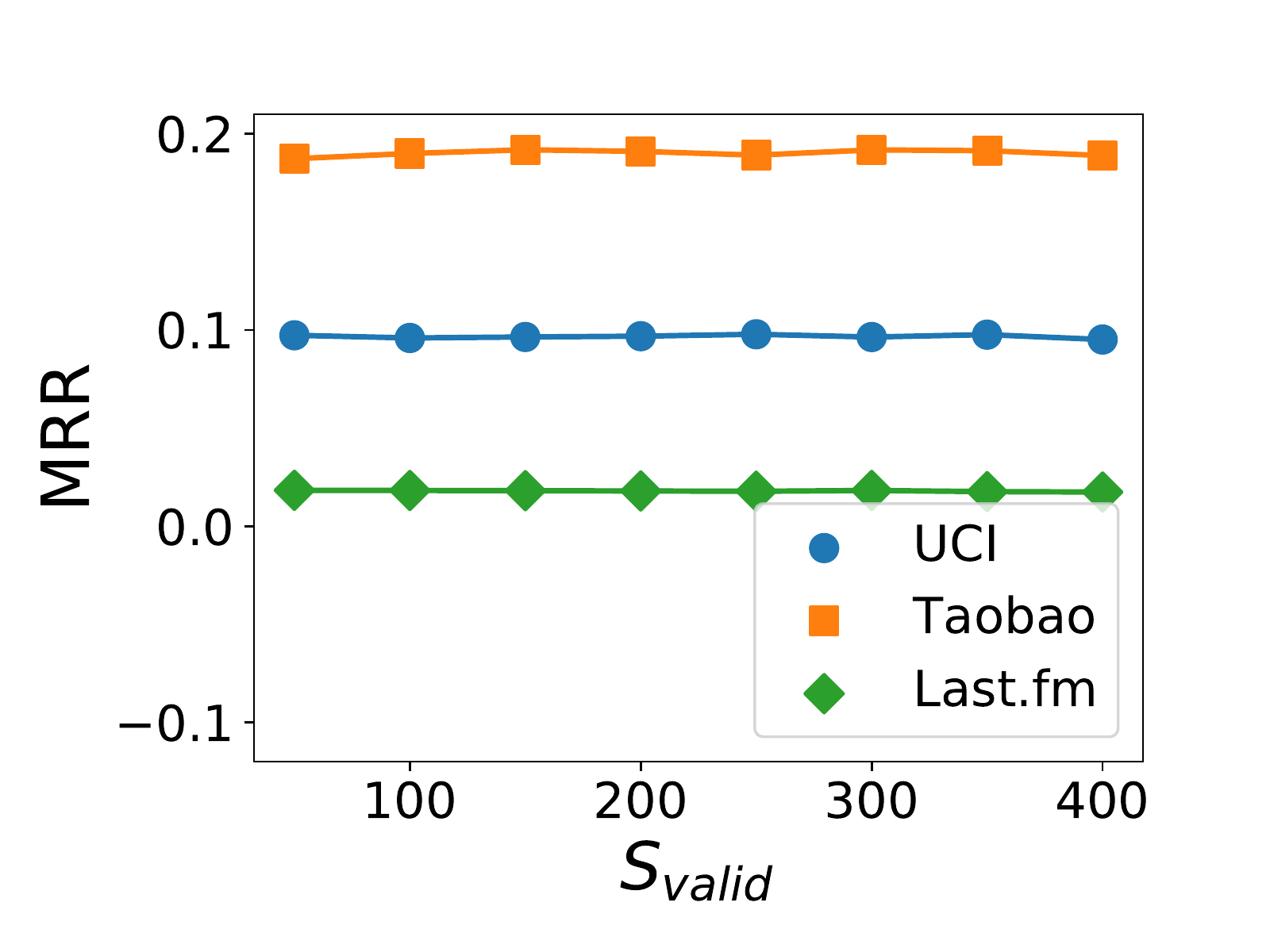}
    \end{minipage}
  }
  \subfigure[Impact of $\mu$]{
    \begin{minipage}[]{0.17\linewidth}
  \includegraphics[width=1.0\linewidth]{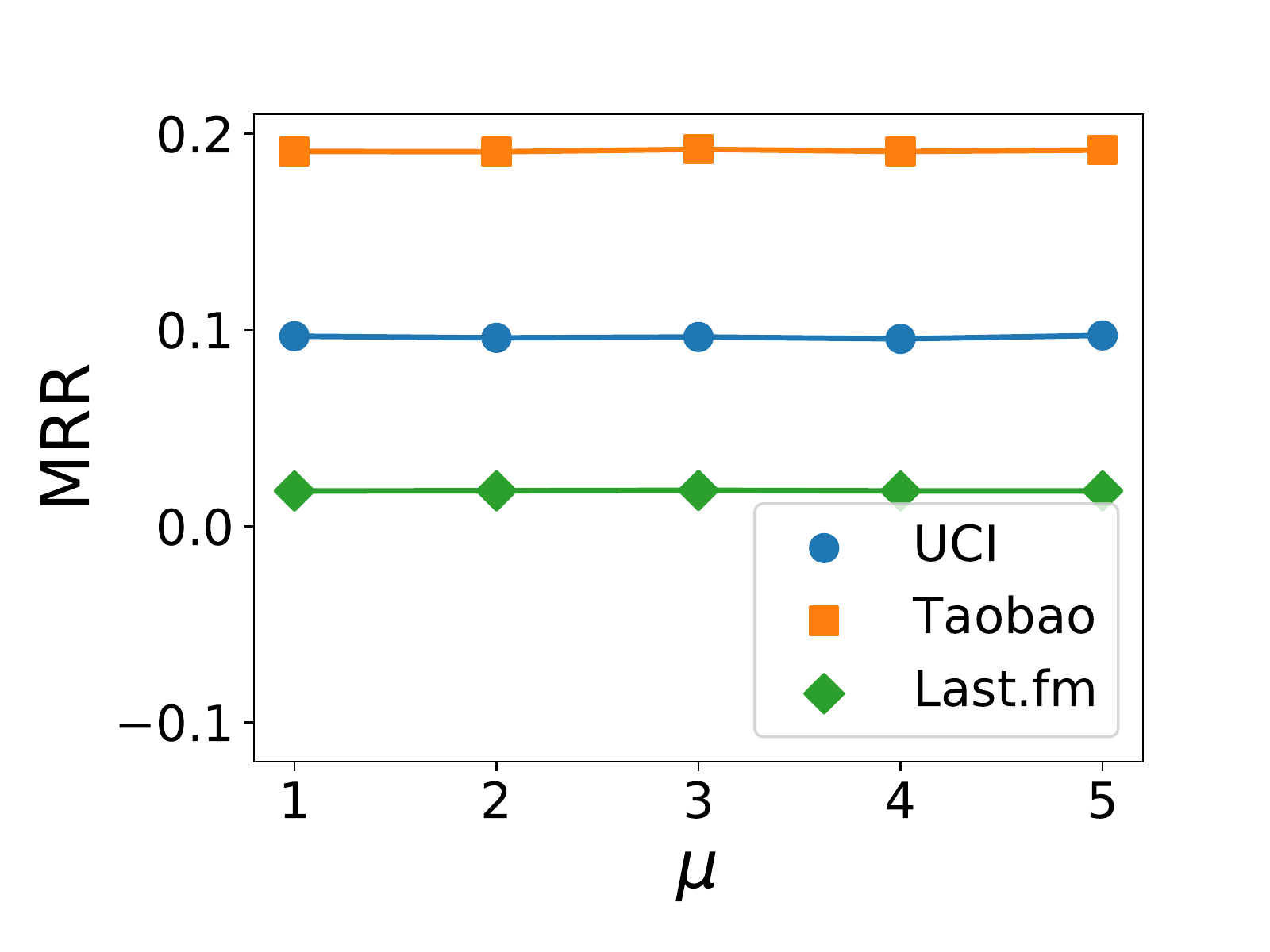}
    \end{minipage}
  }
  \subfigure[\change{Impact of $S_{batch}$}]{
    \begin{minipage}[]{0.17\linewidth}
  \includegraphics[width=1.0\linewidth]{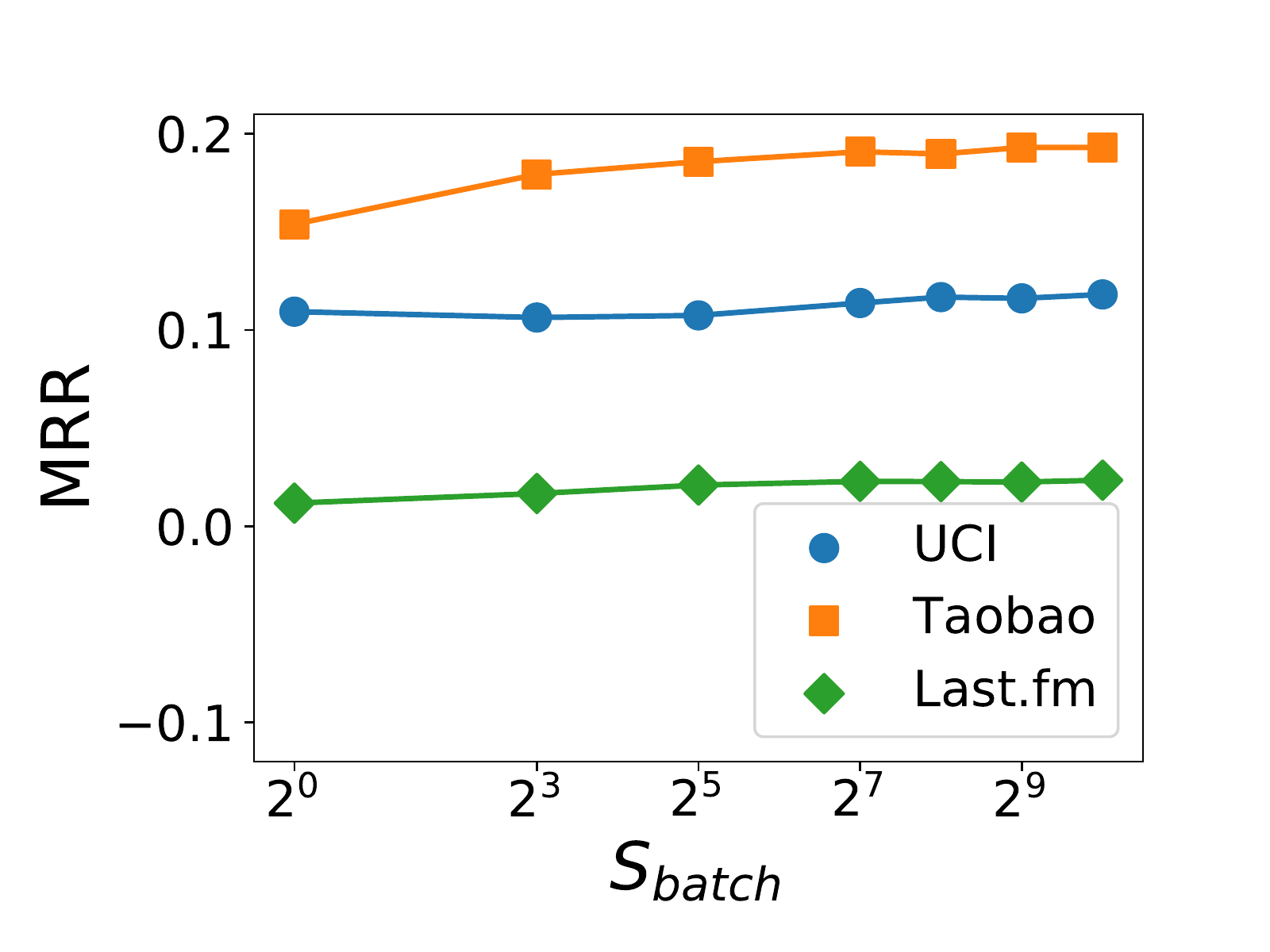}
    \end{minipage}
  }
  \caption{The experimental results of parameter sensitivity.}
  \label{fig:parameter_sensitivity}
  \vspace{-5pt}
\end{figure*}


\subsection{Ablation Study}
In this subsection, we analyze the effects of different components of SUPA.

\subsubsection{Contribution of Different Kinds of Losses}
In this part, we analyze the effectiveness of the three different kinds of losses in SUPA, namely $\mathcal{L}_{inter}$, $\mathcal{L}_{prop}$ and $\mathcal{L}_{neg}$. More precisely, we consider all the usage combinations of these losses, and thus six variants are obtained.
Specifically, SUPA$_{\mathcal{L}_{inter}}$, SUPA$_{\mathcal{L}_{prop}}$, and SUPA$_{\mathcal{L}_{neg}}$ are the variants that use only $\mathcal{L}_{inter}$, $\mathcal{L}_{prop}$ and $\mathcal{L}_{neg}$, respectively, while SUPA$_{w/o \mathcal{L}_{inter}}$, SUPA$_{w/o \mathcal{L}_{prop}}$, and SUPA$_{w/o \mathcal{L}_{neg}}$ are the variants without $\mathcal{L}_{inter}$, $\mathcal{L}_{prop}$ and $\mathcal{L}_{neg}$, respectively.

\enlargethispage{2em}
We report the experimental results in Table~\ref{tab:losses}. As is shown, in general, SUPA has the best performances compared to the variants, and we attribute it to the careful design of the three kinds of losses. We also notice that $\mathcal{L}_{prop}$ is the most important since it enables SUPA to model high-order relationships between nodes, while $\mathcal{L}_{neg}$ is also necessary to avoid model over-smoothing.
Besides, in the MovieLens dataset, SUPA$_{\mathcal{L}_{inter}}$ is better than SUPA and other variants. This is because that MovieLens is dense, and thus the interaction information among neighbors is enough for training, while incorporating high-order relationships and negative sampling may introduce noises.

\subsubsection{Benefits of Modeling Multiplex Heterogeneity and Streaming Dynamics}
Specifically, to show the benefits of modeling multiplex heterogeneity, we consider three variants:
1) SUPA$_{sn}$ in which a same $\alpha_o$ is used for different node types in the node-type specific updater,
2) SUPA$_{se}$ in which the relation-specific context embedding is replaced with a single context embedding,
and 3) SUPA$_{s}$ in which all the heterogeneity related components are removed.
To demonstrate the benefit of modeling the streaming dynamics, we also choose three variants:
1) SUPA$_{nf}$ in which the short-term memory of nodes is removed,
2) SUPA$_{nd}$ in which the decreasing function $g(\cdot)$ and the filter function $h(\cdot)$ are removed when propagating interaction information to the influenced graph,
and 3) SUPA$_{nt}$ in which all the time related components are removed.
Then we conduct link prediction to evaluate SUPA with these variants.

The experimental results on the Taobao and Kuaishou datasets are reported in Table~\ref{tab:benefit}. Notice that the similar results are obtained in other datasets.
As we can see, SUPA has the best performance compared to all the variants, which shows that incorporating the multiplex heterogeneity and streaming dynamic indeed enhances node representation learning. Particularly, SUPA$_{s}$ and SUPA$_{nt}$ show the worst results in their group since they fail to benefit from semantic and temporal information, respectively.

\subsubsection{\change{Effectiveness of the InsLearn training flow}}
\enlargethispage{2em}
\change{To investigate the effectiveness of InsLearn, we implement SUPA$_{w/o Ins}$ to compare InsLearn with conventional training workflow. Specifically, SUPA$_{w/o Ins}$ is trained without the batch iteration and validation mechanism, and scans the whole edge set for multiple times to learn optimal parameters.}

\change{According to Table~\ref{tab:losses}, SUPA significantly outperforms SUPA$_{w/o Ins}$ in most cases. SUPA$_{w/o Ins}$ obtains comparable results to SUPA on Amazon. The reason is that Amazon is not a dynamic dataset, and thus the performance gain by InsLearn is not obvious. Note that SUPA is trained and validated within a single pass and is more efficient than SUPA$_{w/o Ins}$.}

\begin{figure}[t]
  \scriptsize
  \centering
  \subfigure[node2vec ($\bar{d} = 1925$)]{
    \begin{minipage}[]{0.39\linewidth}
    \includegraphics[width=1.0\linewidth]{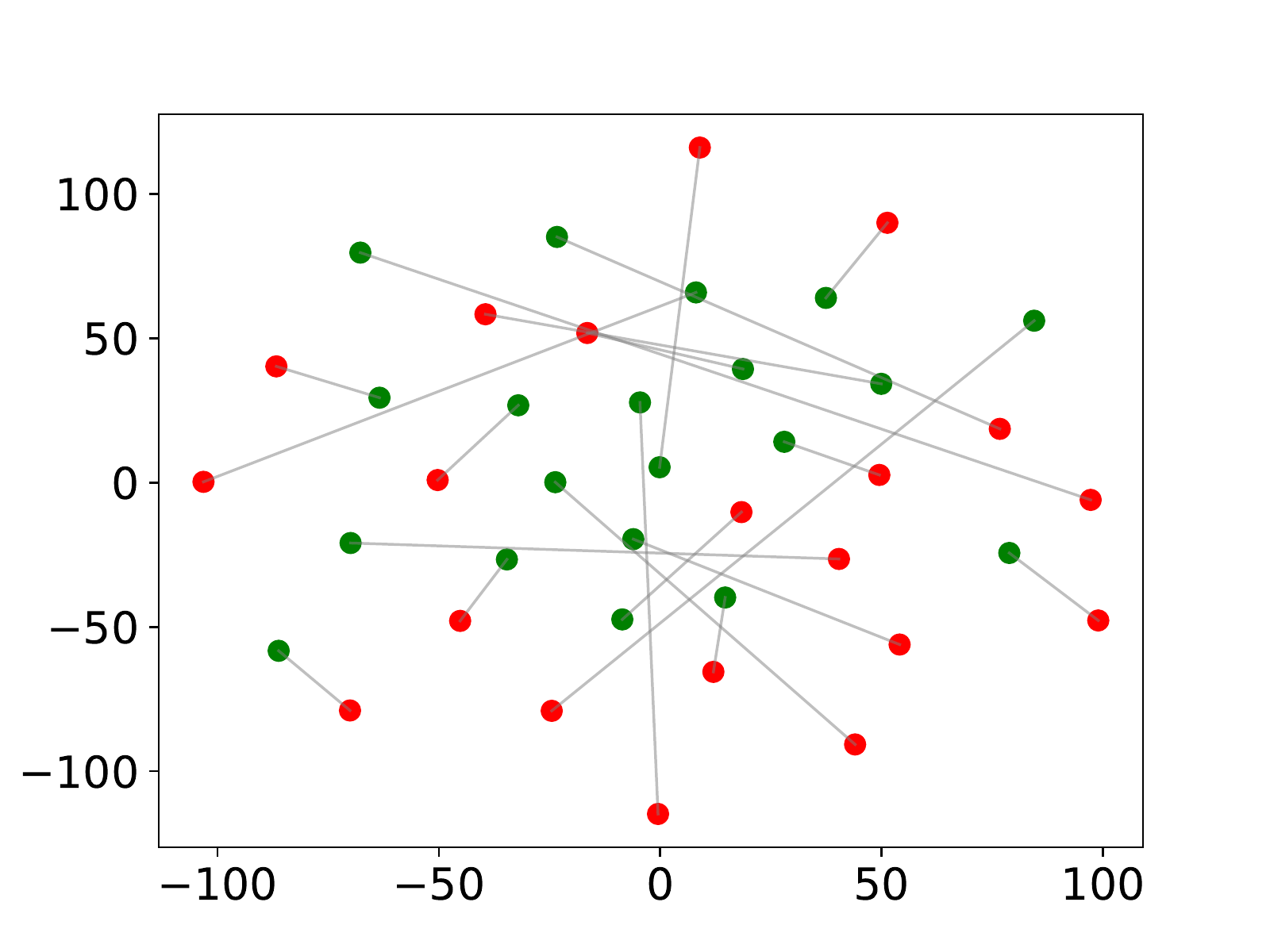}
    \end{minipage}
  }
  \subfigure[GATNE ($\bar{d} = 1848$)]{
    \begin{minipage}[]{0.39\linewidth}
  \includegraphics[width=1.0\linewidth]{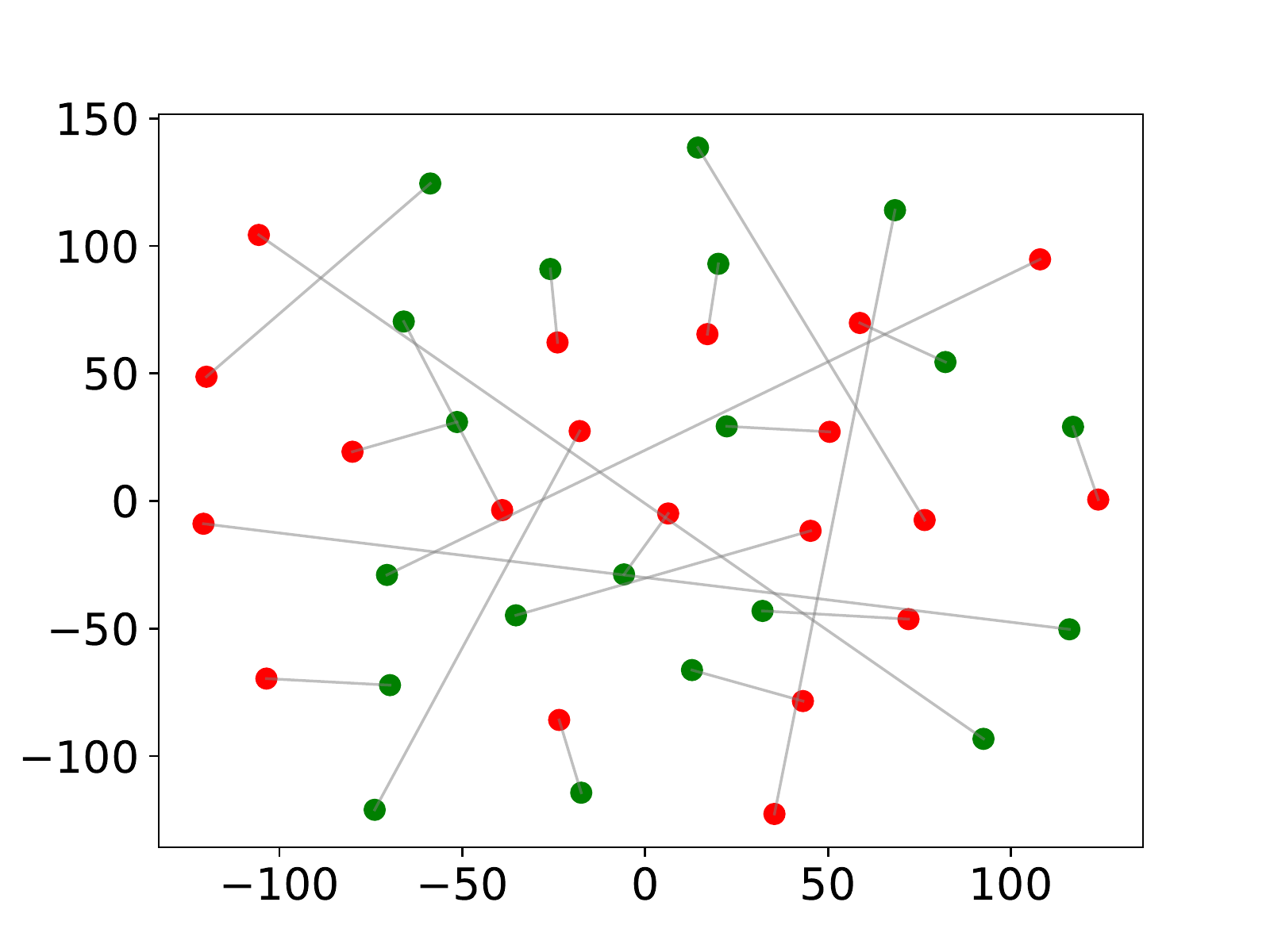}
    \end{minipage}
  }
  \subfigure[LightGCN ($\bar{d} = 2184$)]{
    \begin{minipage}[]{0.39\linewidth}
  \includegraphics[width=1.0\linewidth]{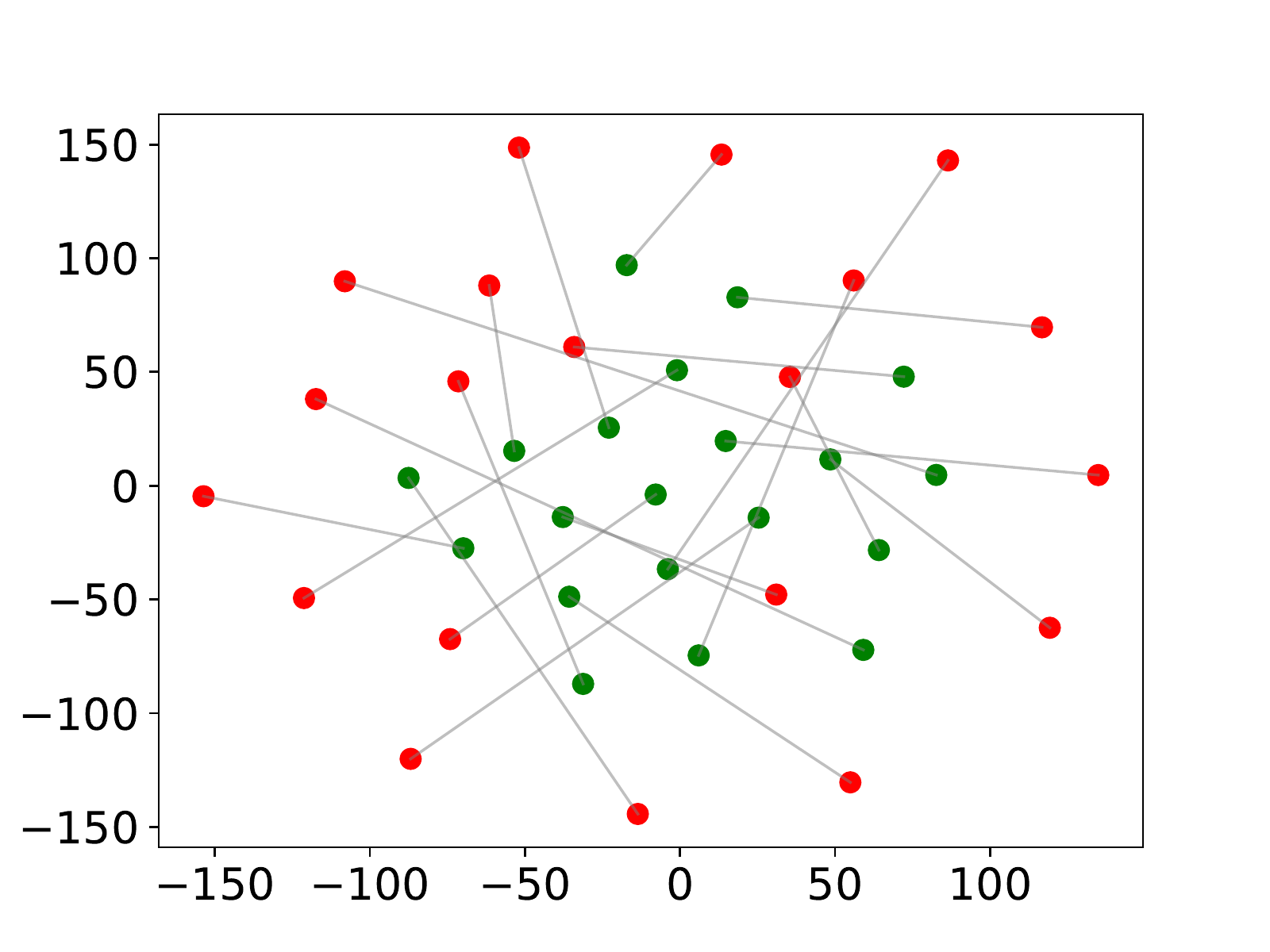}
    \end{minipage}
  }
  \subfigure[MB-GMN ($\bar{d} = 1946$)]{
    \begin{minipage}[]{0.39\linewidth}
  \includegraphics[width=1.0\linewidth]{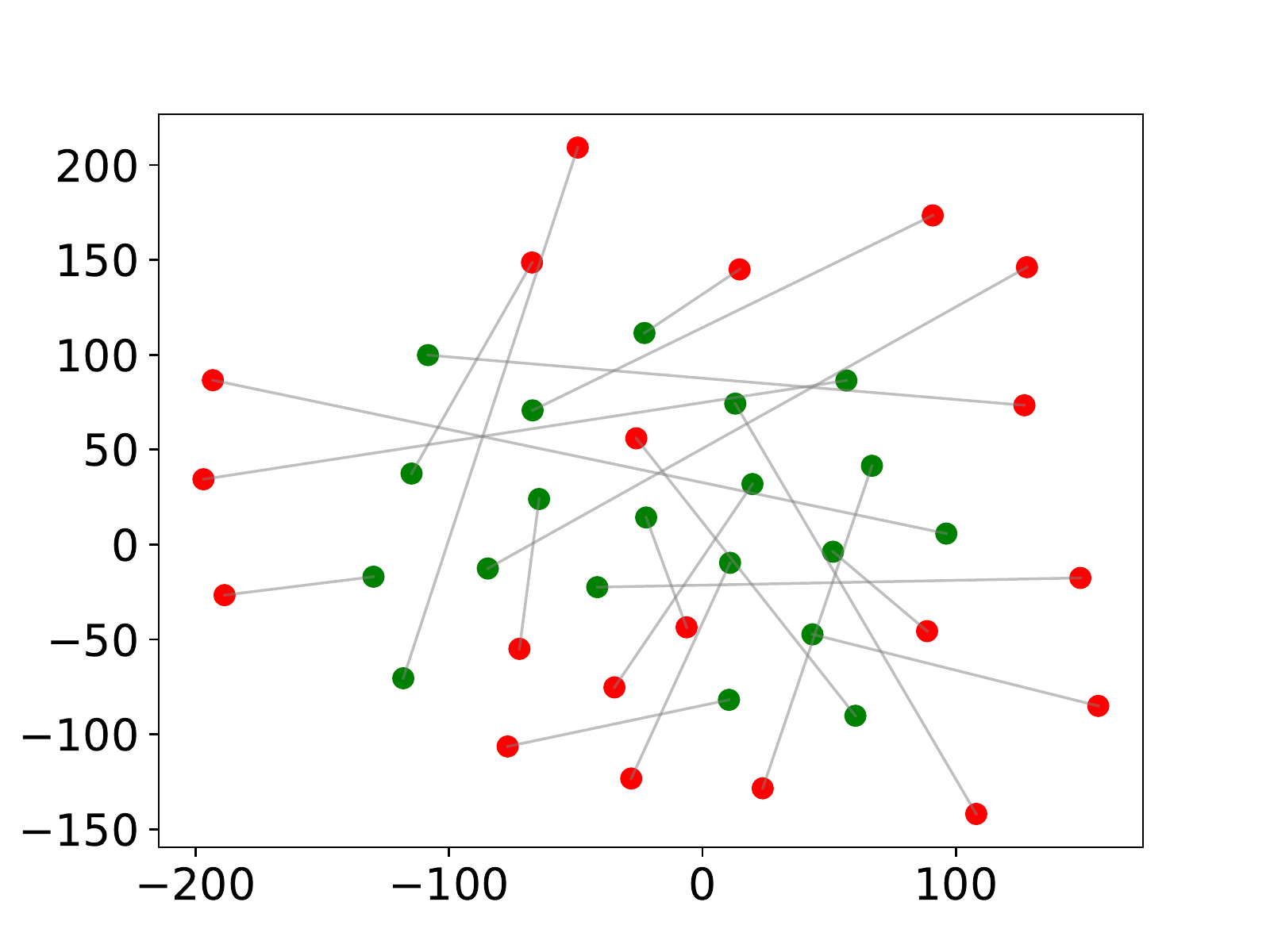}
    \end{minipage}
  }
  \subfigure[EvolveGCN ($\bar{d} = 2448$)]{
    \begin{minipage}[]{0.39\linewidth}
  \includegraphics[width=1.0\linewidth]{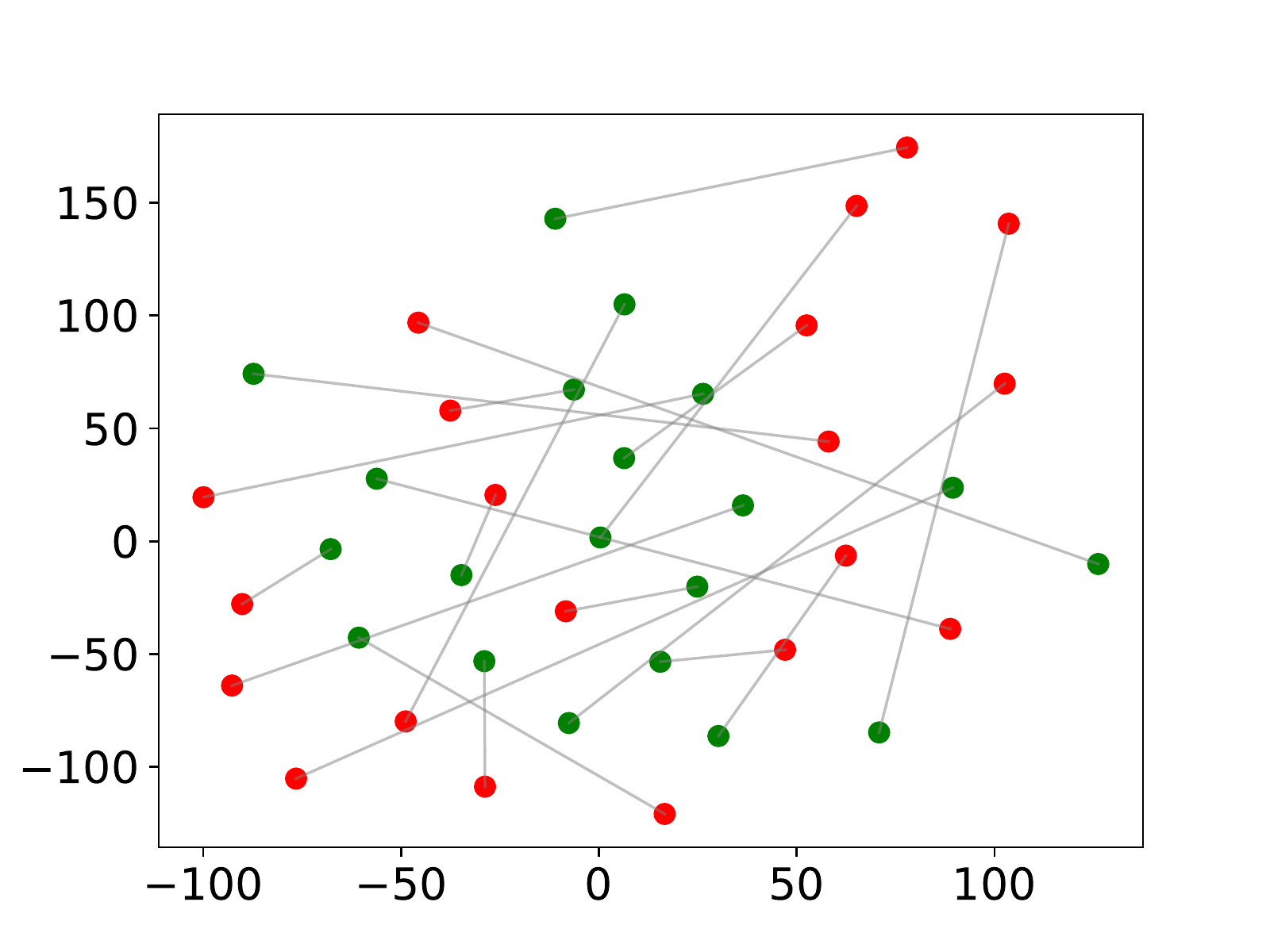}
    \end{minipage}
  }
  \subfigure[SUPA ($\bar{d} = 1450$)]{
    \begin{minipage}[]{0.39\linewidth}
    \includegraphics[width=1.0\linewidth]{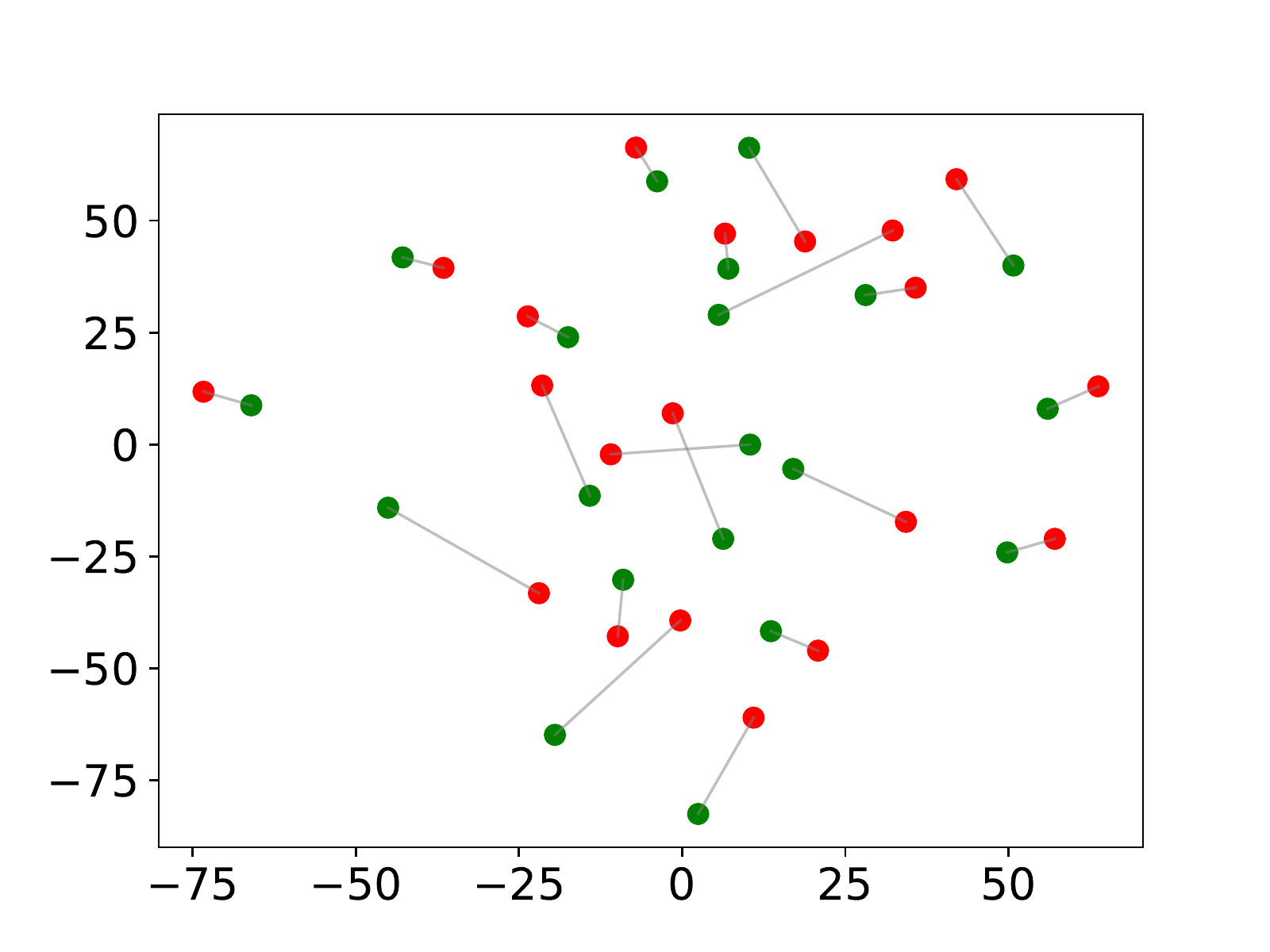}
    \end{minipage}
  }
  \caption{Embedding visualization of user-item pairs in Taobao. Users and items are represented as red and green nodes, respectively. The user-item pairs are linked by gray lines. The x-axis and y-axis denote the coordinates.}
  \label{fig:visual_link}
\end{figure}


\subsection{\change{Scalability}}
\change{In this subsection, we investigate the scalability of SUPA in terms of fast changing data. Specifically, we adjust the batch size of the training set (i.e., $S_{batch}$) and test the average running time (i.e., the average re-training time given $S_{batch}$ new edges) among all batches.
Figure~\ref{fig:scalability} shows the running time and the recommendation performance of SUPA on MovieLens and similar tendencies are observed in other datasets. Note that the horizontal axis is in log scale. It is observed that the running time is linear to $S_{batch}$ while the recommendation performance remains stable. Specially, when $S_{batch} = 2^{15}$, the average running time is 2.811s, i.e., SUPA is able to deal with 11,657 edges per second using a single GTX 1080Ti GPU, which shows the scalability of SUPA. To deal with larger dynamic graph, one can use multiple GPU to train SUPA since the update procedure of SUPA is localized.}

\subsection{Parameter Sensitivity}
In this subsection, we analyze the sensitivity of hyper-parameters in SUPA. Specifically, for the GNN model, we examine the impact of the embedding size $d$, the number of walk $k$, the walk length $l$, the number of negative sample $N_{neg}$ and the filter threshold $\tau$.
For the single-pass training workflow,
we consider the influence of the maximum iteration number $N_{iter}$, the validation interval $I_{valid}$, the size of validation set $S_{valid}$, the patience for early stopping $\mu$ \change{and the size of each training batch $S_{batch}$}.
\enlargethispage{2em}
The experiments are conducted on the UCI, Last.fm and Taobao datasets, and similar tendencies are observed in other datasets. The results are illustrated in Figure~\ref{fig:parameter_sensitivity}.
For the hyper-parameters in the GNN model, we observe that the performance has little improvement after $d$ is above 128 (see Figure~\ref{fig:parameter_sensitivity}a). Besides, the parameters $k$ and $l$ have different impact to the performance on different datasets (see Figure~\ref{fig:parameter_sensitivity}b and \ref{fig:parameter_sensitivity}c), and thus need to be carefully chosen. Finally, for $N_{neg}$ and $\tau$, empirically setting $N_{neg}=5$ and $g(\tau)=0.3$ can always give comparable results (see Figure~\ref{fig:parameter_sensitivity}d and \ref{fig:parameter_sensitivity}e).

For the hyper-parameters in the training workflow,
we find that a smaller $N_{iter}$ is preferred on a dataset with more edges (see Figure~\ref{fig:parameter_sensitivity}f) to avoid over-fitting. In contrast, the other parameters are less sensitive (see Figure~\ref{fig:parameter_sensitivity}g, \ref{fig:parameter_sensitivity}h and \ref{fig:parameter_sensitivity}i) so we can choose larger $I_{valid}$, smaller $\mu$ and $S_{valid}$ to accelerate the training procedure.
\change{Besides, a small $S_{batch}$ (i.e., $S_{batch} < 32$) can result in over-fitting and thus has negative impact to the performance (See Figure~\ref{fig:parameter_sensitivity}j). SUPA is not sensitive to $S_{batch}$ when $S_{batch} \ge 32$.}

\subsection{Visualization}
In addition to quantitative evaluations of SUPA, we conduct a qualitative assessment by visualizing the embeddings generated by SUPA and several baseline methods in the link prediction task.
Specifically, we randomly select 20 user-item pairs from the testing set of the Taobao dataset, and then use t-SNE~\cite{van2008visualizing} to project the embeddings of these nodes into a 2-dimensional space.
For fair comparation, we also repeat the procedure for 100 times and calculate the average sum $\bar{d}$ of distances between nodes after t-SNE projection.

Figure~\ref{fig:visual_link} shows the visualization results of node2vec, GATNE, LightGCN, MB-GMN, EvolveGCN, and SUPA.
As we can see, SUPA obtains more separate short lines than the baseline methods, which indicates that SUPA learns similar embeddings for nodes in each user-item pair in the test set. Thus, SUPA can make better recommendations than other methods based on the similarity between the embeddings of true node pairs.

\section{Related Work}
In this section, we review related work relevant to our study, including multi-relation modeling for recommendation and graph modeling for recommendations.
\label{sec:related_work}
\subsection{Multi-relation Modeling for Recommendation}
Multi-relation modeling is proposed to enhance representation learning for recommendations with the consideration of various relations from the side information of users and items~\cite{huangrecent,liu2020heterogeneous,xia2021knowledge}.
There exist mainly three categories of  work on multi-relation modeling.
The first category of studies propose to incorporate social relations of users to alleviate the data sparsity issue and thus boost the prediction performance of users' preferences by considering the influence between different users~\cite{chen2019efficient,huang2021knowledge,wu2019neural,fan2019graph,yu2018walkranker}.
The second category of multi-relation modeling leverage external knowledge graph information to supplement the interaction learning between users and items by constructing various structural relations between different entities~\cite{wang2019knowledge,wang2019kgat,zhu2020knowledge}.
The third category is multi-behavior modeling, which utilizes multiple user-item feedback to enhance recommendation on target behaviors. A research line of works approaches this task in a multi-task manner, i.e., considers recommendation for different types of behavior as different tasks and share parameters across tasks for joint learning~\cite{chen2020efficient,gao2019learning,tang2016empirical}. Another paradigm of multi-behavior models uses the multiplex user behaviors as assistant to comprehend user intents and presents negative sampling based solutions to make recommendations~\cite{ding2018improving,loni2016bayesian,jin2020multi,xia2021graph}.

In this paper, we focus on multi-behavior modeling and incorporate temporal information to characterize the drift of user interest and the propagation of behavior influence.
%

\subsection{Graph Modeling for Recommendation}
\todo{Graph modeling for recommendation leverages graph embedding methods to learn representations of users, items as well as auxiliary knowledge graph entities (if there is any), and makes recommendations using the link prediction task based on the learned representations.}

Early graph embedding methods focus on topological structure to learn representations~\cite{perozzi2014deepwalk, node2vec-kdd2016,tang2015line,hamilton2017inductive,velivckovic2017graph,he2020lightgcn}.
To incorporate semantic information,  metapath-based patterns are proposed to perform heterogeneity-aware graph sampling and aggregation~\cite{zhang2019heterogeneous,wang2019neural,zheng2020price,chen2019collaborative,dong2017metapath2vec,wang2019heterogeneous,fu2020magnn}.
Furthermore, a group of graph embedding studies notices that real-world graphs are more complicated where exist multiple types of relations between nodes~\cite{liu2017principled,qu2017attention}, and starts to model the multiplex heterogeneity of graphs by learning a common embedding and several additional edge-type specific embeddings for each node~\cite{zhang2018scalable,shi2018mvn2vec,cen2019representation,tiankai2022hybridgnn}.

Another research line of works is designed to deal with dynamic graphs and the fundamental idea is to extend static embedding methods to model the temporal variation. Specifically, dynamic modelling includes snapshot-by-snapshot update~\cite{li2017attributed,wang2020dynamic,yin2019dhne,bian2019network,fard2019relationship}, temporal random walk sampling~\cite{nguyen2018continuous,yu2018netwalk}, temporal regularization or encoding~\cite{goyal2018dyngem,zhou2018dynamic,ruan2021efficient,xu2020inductive}, and GNN-based recurrent architecture or temporal attention~\cite{pareja2020evolvegcn,ma2020streaming,Xue2020DyHATR,yang2020dynamic,sajadmanesh2019continuous,peng2021lime}.

However, most of the above works fail to model multiplex heterogeneity and streaming dynamics simultaneously. Furthermore, as far as we know, no techniques are proposed to deal with neighborhood disturbance yet.

\section{Conclusion}
\label{sec:conclusion}
In this paper, we analyzed the neighborhood disturbance phenomenon in dynamic graphs, and formally defined the problem of DMHG instant representation learning. Then we proposed SUPA, a novel GNN model to solve it. Specifically, SUPA developed a sample-update-propagate architecture to capture the multiplex heterogeneity and streaming dynamics simultaneously.
Besides, we designed InsLearn, a single-pass training workflow to train SUPA in an incremental manner.
Through comprehensive experiments on six real-world datasets, we empirically showed the superiority and the generalization ability of SUPA.
\change{In the future, SUPA will be developed to explore the constrains on the edge type sets of multiplex metapath schemas and compute the set of multiplex metapath schemas automatically.}

\section*{Acknowledgments}
This work is supported in part by the National Natural Science Foundation of China (No.~61872207) and Kuaishou Inc.
Chaokun Wang is the corresponding author.

\bibliographystyle{IEEEtran}
\bibliography{SUPA}

\end{document}